\documentclass[aps,twocolumn,superscriptaddress,floatfix]{revtex4-1}
\usepackage{graphicx,amsfonts,amssymb,amsmath,mathrsfs,hyperref,bm,dsfont,lipsum,esvect}
\usepackage{relsize}
\usepackage{comment}
\usepackage{xcolor}
\usepackage{soul}

\newif\ifhyper
\hypertrue 
\ifhyper
\hypersetup{
   citecolor = {red},
   colorlinks = {true}, 
   linkcolor = {blue},
   urlcolor = {blue} 
}
\fi

\def\be{\begin{equation}}
\def\ee{\end{equation}}
\def\bea{\begin{eqnarray}}
\def\eea{\end{eqnarray}}

\newcommand{\ket}[1]{|#1\rangle}
\newcommand{\bra}[1]{\langle #1|}

\newcommand{\expectval}[1]{\langle #1\rangle}

\begin{document}

\title{Kitaev honeycomb antiferromagnet in a field:  \\ quantum phase diagram for general spin}
\author{Saeed S. Jahromi}
\email{saeed.jahromi@iasbs.ac.ir}	
\affiliation{Donostia International Physics Center, Paseo Manuel de Lardizabal 4, E-20018 San Sebasti\'an, Spain}
\affiliation{Department of Physics, Institute for Advanced Studies in Basic Sciences (IASBS), Zanjan 45137-66731, Iran}
\affiliation{Multiverse Computing, Paseo de Miram\'on 170, E-20014 San Sebasti\'an, Spain}
\author{Max H\"ormann}
\affiliation{Department of Physics, Staudtstra{\ss}e 7, Friedrich-Alexander-Universit\"at Erlangen-N\"urnberg (FAU), D-91058 Erlangen, Germany}
\author{Patrick Adelhardt}
\affiliation{Department of Physics, Staudtstra{\ss}e 7, Friedrich-Alexander-Universit\"at Erlangen-N\"urnberg (FAU), D-91058 Erlangen, Germany}
\author{Sebastian Fey}
\affiliation{Department of Physics, Staudtstra{\ss}e 7, Friedrich-Alexander-Universit\"at Erlangen-N\"urnberg (FAU), D-91058 Erlangen, Germany}
\author{Hooman Karamnejad}
\affiliation{Department of Physics, Institute for Advanced Studies in Basic Sciences (IASBS), Zanjan 45137-66731, Iran}
\author{Rom\'an Or\'us}
\affiliation{Donostia International Physics Center, Paseo Manuel de Lardizabal 4, E-20018 San Sebasti\'an, Spain}
\affiliation{Ikerbasque Foundation for Science, Maria Diaz de Haro 3, E-48013 Bilbao, Spain}
\affiliation{Multiverse Computing, Paseo de Miram\'on 170, E-20014 San Sebasti\'an, Spain}
\author{Kai Phillip Schmidt}
\affiliation{Department of Physics, Staudtstra{\ss}e 7, Friedrich-Alexander-Universit\"at Erlangen-N\"urnberg (FAU), D-91058 Erlangen, Germany}

\begin{abstract}
We combine tensor-network approaches and high-order linked-cluster expansions to investigate the quantum phase diagram of the antiferromagnetic Kitaev's honeycomb model in a magnetic field for general spin values. For the pure Kitaev model, tensor network calculations confirm the absence of fluxes and spin-spin correlations beyond nearest neighbor in the ground state, but signal a breaking of the discrete orientational symmetry for $S\in\{1,3/2,2\}$ inline with the semiclassical limit. An intermediate region between Kitaev phases and the high-field polarized phase is demonstrated for all considered spin values. In this intermediate region the tensor network results display a sequence of potential phases whose number increases with the spin value. Each of these can be characterized by distinct local magnetization patterns while the total magnetization increases smoothly as a function of the field. The analysis of the high-field zero-momentum gap and the associated spectral weight of the polarized phase for general spin $S$ obtained by linked-cluster expansions is consistent with an unconventional quantum critical breakdown of the high-field polarized phase in accordance with the presence of exotic physics at intermediate Kitaev couplings. 
\end{abstract}

\maketitle

\section{Introduction} 
\label{sec:intro}
Quantum spin liquids (QSLs) are exotic phases of matter which remain disordered for all temperatures and host a variety of fascinating physical properties such as, emergent gauge structures \cite{Balents2010, Savary2017}. This includes two- and three-dimensional gapped QSLs with intrinsic topological order displaying highly entangled ground states and unconventional anyonic statistics \cite{Leinaas77p1,Wilczek_1982,Levin2006,Jahromi2017,Capponi2014, Jahromi2016}, which are at the heart of topological quantum computation \cite{Kitaev2006,Nayak2008, Kitaev2003, Kitaev2006,Levin2005}, as well as 3d fracton topological order \cite{Chamon_2005,Bravyi_2011,Haah_2011}, whose robustness is believed to be relevant for topologically protected quantum memories. In contrast, their gapless counterparts display distinct prominent features like emergent (3+1)d quantum electrodynamics in the Coulomb QSL \cite{Hermele_2014,Shannon_2012,Roechner_2016}, but also algebraic Dirac spin liquids \cite{Picot2016,Liao2017,Xie2014,He2017} as well as spin-Bose metals exist in several 2d frustrated quantum magnets \cite{Motrunich_2005,Lee_2005,Sheng_2009,Yang_2010}. 

Kitaev's honeycomb model \cite{Kitaev2006} is one of the most prominent systems in this context, since its spin-1/2 version is exactly solvable in terms of Majorana fermions and realizes gapped and gapless QSLs in the ground-state phase diagram. These QSLs originate from frustrated bond-dependent Ising interactions. 
The fate of the gapless Kitaev QSL in the presence of a field along the (1,1,1)-direction is qualitatively different for ferro- and antiferromagnetic Kitaev interactions \cite{Zhu_2018, Gohlke_2018, Hickey2019}. In the ferromagnetic case, numerical investigations for isotropic Kitaev couplings consistently show a direct transition between the gapped Kitaev QSL with non-Abelian topological order at finite fields and the high-field polarized phase. In contrast, for antiferromagnetic Kitaev interactions, several numerical approaches show the presence of an intermediate phase \cite{Zhu_2018,Gohlke_2018,Hickey2019,Hickey_2020}. This intermediate phase is claimed to be a gapless QSL with an emergent U(1) gauge structure and therefore (2+1)d quantum electrodynamics \cite{Hickey2019}. Furthermore, calculations based on exact diagonalization (ED) \cite{Hickey_2020,Hickey_2021}, density matrix renormalization group (DMRG) \cite{Zhu_2020}, and tensor networks (TN) \cite{Lee2020} indicate the presence of a similar intermediate region for the antiferromagnetic spin-one Kitaev model in a field. 

The presence of an intermediate region for $S\in\{1/2,1\}$ immediately calls for an investigation of the antiferromagnetic Kitaev model in a field for general spin values $S$. Indeed, it was conjectured that an intermediate U(1) QSL is present for all $S$ and that the phase transition between the two QSLs at low and intermediate fields is given by an Anderson-Higgs transition \cite{Hickey_2020,zhang_2021}. At the same time the phase transition between the high-field polarized phase and the intermediate Kitaev region is only poorly understood. In this paper, we gain further insights in the ground-state phase diagram for general $S$ by applying complementary TN approaches as well as linked-cluster expansions in the thermodynamic limit.

The paper is organized as follows. In Sec.~\ref{sec:model} we introduce the Kitaev honeycomb model in the presence of the magnetic field for general spin $S$. Next in Sec.~\ref{sec:method}, our numerical simulation techniques, i.e., the TN method as well as the linked-cluster expansions are briefly reviewed. We provide the numerical results for the phase diagram of the model for different spin values and their interpretation in Sec.~\ref{sec:phase_diag}. Finally, Sec.~\ref{sec:conclude} is devoted to conclusion and outlook for future studies.

\section{Model} 
\label{sec:model}
We study the Kitaev's honeycomb model in a uniform magnetic field along the (1,1,1)-direction, so that the Hamiltonian reads
\begin{align}
 \mathcal{H}_{\rm KIF}=J\sum_{{\alpha\text{-links} \atop \alpha=x,y,z }\atop <i,j>}S_i^\alpha S_j^\alpha+h\sum\limits_{i}\left(S_i^x+S_i^y+S_i^z\right), \label{eq:kitaev_hamiltonian_111_field}
\end{align}
where $S_i^\alpha$ are spin-$S$ operators with $[S_i^\alpha,S_i^\beta]=i\epsilon_{\alpha\beta\gamma}S_i^\gamma$, $J>0$ (antiferromagnetic), and we choose $h$ to be positive. The pure Kitaev model for $h=0$ has a conserved quantum 
number per plaquette $p$ for general $S$ given by the eigenvalues $\pm 1$ of the plaquette operators \mbox{$W_p=e^{{\rm i}\pi\left(S_1^x+S_2^y+S_3^z+S_4^x+S_5^y+S_6^z\right)}$}, which is the basis for the exact solution for $S=1/2$ yielding a flux-free (i.e., $W_p=+1$ for all $p$) and a gapless QSL ground state \cite{Kitaev2006}. For $S>1/2$, the system is not exactly solvable anymore,  but it is known that even in the classical limit, the pure Kitaev model does not order magnetically due to the strong frustration \cite{Baskaran2007}. In the semi-classical limit, the ground state remains flux free but it is a gapped $\mathbb{Z}_2$ QSL with vanishing spin-spin correlations beyond nearest-neighbors and a breaking of orientational symmetry \cite{Baskaran2008b,Rousochatzakis_2018}. The semi-classical approximation is suggested to be valid for $S\gtrsim 3/2$. For smaller spin values, however, it is believed that tunneling between different dimer patterns leads to its breakdown \cite{Rousochatzakis_2018}. Moreover, DMRG calculations for $S=1$ \cite{Zhu_2020} and $S=3/2$ \cite{jin_2021} predict a gapless QSL ground state similar to $S=1/2$ while TN calculations \cite{Lee2020} predict a gapped QSL for $S=1$ inline with the semi-classical findings \cite{Rousochatzakis_2018}. For finite fields $h>0$ and $S\in\{1/2,1\}$ there is convincing evidence by ED \cite{Hickey2019,Hickey_2020,Hickey_2021} and DMRG \cite{Zhu_2018,Gohlke_2018,Zhu_2020} for the presence of an intermediate region in the phase diagram between the Kitaev phase at small (but finite) fields and the high-field polarized phase, which is also gapped but topologically trivial. Numerical simulations in Refs.~\cite{Zhu_2018,Gohlke_2018,Hickey2019,Hickey_2020,Hickey_2021} indicate that this intermediate phase is gapless for both $S=1/2$ and $S=1$. 

\begin{figure}[t]
\centerline{\includegraphics[width=\columnwidth]{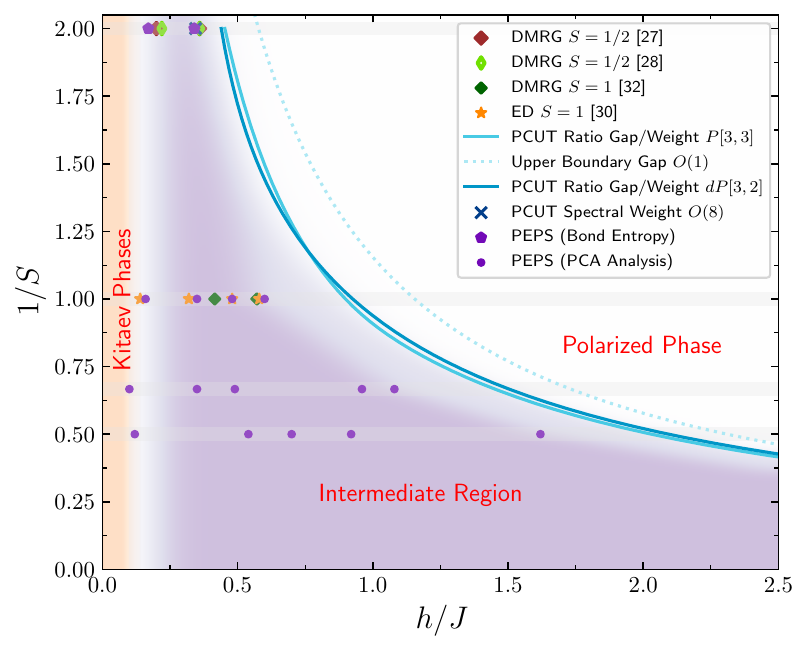}}
\caption{{\bf Phase diagram of the antiferromagnetic spin-$S$ Kitaev's honeycomb model in a (1,1,1)-field as a function of $h/J$ and $1/S$.} Filled circles denote the iPEPS data and correspond to the $h/J$ boundaries between different phases and sub-regions shown in Fig.~\ref{fig3}, while lines originate from the high-field series expansion. Solid lines correspond to Pad\'{e} (DlogPad\'{e}) approximant $P[L,M]$ ($dP[L,M]$) of the ratio gap $\Delta^S$ over spectral weight $A^S_{\mathrm{Gap}}$ from pCUT (see Appendix~\ref{apndx:pcut-extrapl} for definition of respective approximants). The DMRG and ED benchmark data were obtained from Refs.~\cite{Gohlke_2018,Zhu_2018,Zhu_2020} and \cite{Hickey_2020}, respectively.}
\label{Fig:phase-diag}
\end{figure}

\begin{figure*}[t]
	\centering
	\begin{minipage}{\textwidth}
		\centering
		\includegraphics[width=\textwidth]{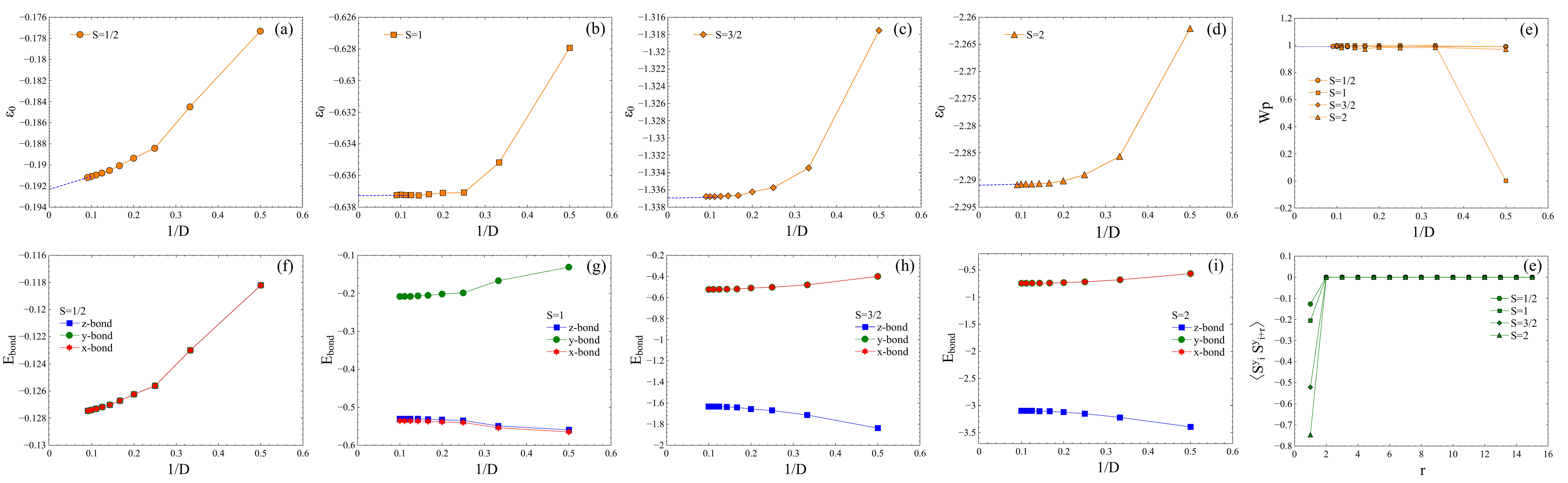}
	\end{minipage}
	\caption{{\bf Tensor Network results for the pure spin-$S$ Kitaev model with $S\in\{1/2,1,3/2,2\}$ obtained with simple update algorithm and corner transfer matrix renormalization group.} (a)-(d) Ground-state energy per-site, $\varepsilon_0$. (f) Plaquette operator $W_p$.  (f)-(i) Bond energy $E_{\rm bond}=J_\alpha\expectval{S_i^\alpha S_{i+1}^\alpha}$ ($\alpha$-links) as a function of $1/D$. And (j) Spin-spin correlation functions $\expectval{S_i^yS_{i+r}^y}$, where  $r$ is the distance from a green link along the zigzag chain connected to that link. For other spin flavors we obtain similar correlation results.}
\label{fig2}
\end{figure*}

\section{Methods}
\label{sec:method}
\subsection{Tensor Network} Tensor-Network (TN) methods \cite{Orus2014,Orus2014a,Orus2019,Ran2017,Biamonte2017,Verstraete2008} provide efficient representations for ground states of local Hamiltonians based on their entanglement structure \cite{Orus2014}. The projected entangled-pair state (PEPS) method and its infinite version in the thermodynamic limit (iPEPS) \cite{Verstraete2006,Verstraete2008,Orus2014,Orus2014a} have played a major role in the characterization and discovery of many exotic phases, ranging from magnetically ordered states \cite{Orus2009,Corboz2013,Phien2015} to QSLs \cite{Picot2016,Xie2014,Jahromi2021,Jahromi2020} and valence-bond crystals \cite{Iqbal2018,Jahromi2018a}. In particular, the modified version of the iPEPS algorithm, designed for the honeycomb structures \cite{Jahromi2018}, has been shown to be very successful for simulating and characterizing the Kitaev model and its variants, such as the Kitaev-Heisenberg model, in the thermodynamic limit \cite{OsorioIregui2014,Czarnik2019,Jahromi2019}. 

In this study we use different variants of the iPEPS algorithm based on simple update (SU) \cite{Jiang2008,Corboz2010a,Jahromi2019}, and graph-based PEPS algorithm (gPEPS) \cite{Jahromi2019}, in order to simulate the Kitaev model in the presence of a uniform magnetic field for generic spin-$S$ values. Both SU and gPEPS algorithms use mean-field-like environment and local tensor updates based on singular-value decomposition to approximate the local PEPS tensors of the ground state wave function. We simulated the Kitaev model for different $S$ on different unit-cells with $8, 12, 16$ and $32$ sites and different bond dimensions $D$. The maximum achievable dimensions were $D=11$ and $18$ for SU and gPEPS simulations, respectively. Our optimization algorithms in all aforementioned approaches was based on imaginary-time evolution \cite{Orus2008,Orus2014} with $\delta\tau$ starting from $10^{-1}$ down to $10^{-5}$ with $4000$ maximum iterations to ensure both convergence and accuracy.   

Let us further note that we used the corner transfer matrix renormalization group (CTMRG) method \cite{Nishino1996,Orus2009,Corboz2014a,Corboz2010a} for contracting the 2D infinite TN as well as for calculating the variational energies and expectation values of local operators in the SU algorithm. We fixed the boundary dimension to $\chi=D^2$. However, for $D>8$ we used $\chi=64$ due to limitations in computational resources. These values were already sufficient for convergence of the expectation values with respect to boundary dimension. Moreover, we used the mean-field-like environment for calculating the expectation values in the gPEPS technique, thus allowing for a larger achievable bond dimension $D$ \cite{Jahromi2019}. However, one should note that by using the mean-field environments the correlations beyond the nearest neighbors are captured less accurately and the expectation values are no longer variational. In particular, the gPEPS algorithm works well for gapped phases with short range correlations such as the polarized phase, as will be shown in the future sections. However, it should be used with care for simulating the gapless states such as the ground state of the spin-$1/2$ Kitaev model. Extra details about the TN algorithms and their implementation can be found in Appendix~\ref{apndx:tn}.

\subsection{Linked-Cluster Expansions} The method of perturbative continuous unitary transformations (pCUT) \cite{Knetter2000,Knetter2003} maps the Hamiltonian \eqref{eq:kitaev_hamiltonian_111_field} for a general spin value $S$ unitarily to an effective Hamiltonian $\mathcal{H}_{\rm eff}$, which conserves the number of quasi-particles (qp) in the polarized high-field phase where spins point in $(1,1,1)$-direction. This mapping is done perturbatively up to high orders in powers of $J/h$. The quasi-particles in the polarized phase correspond to dressed spin-flip excitations, which are adiabatically connected to localized spin flips above the fully polarized state \mbox{$\ket{\rm ref}\equiv\ket{-S-S\ldots -S}$} in the limit $h\rightarrow\infty$. We take the field term as the unperturbed part $\mathcal{H}_0$, which has an equidistant spectrum for any spin value $S$. The Kitaev interactions then correspond to the perturbation changing the number of spin flips by $N \in \{0,\pm 1,\pm 2\}$. Employing a white-graph expansion \cite{Coester2015, Adelhardt2020, Adelhardt2024} we performed pCUT calculations in the 0qp and 1qp sector either for fixed spin $S\in\{1/2,1\}$ or by keeping the spin $S$ as a general variable. This allows us to determine the ground-state energy per site $\varepsilon^S_0$ (order 9 for $S=1/2$, order 8 for $S=1$, and order 7 for general $S$) and the one-quasi-particle gap $\Delta^S$ (order 9 for $S=1/2$, order 8 for $S=1$, and order 7 for general $S$) as a high-order polynomial in $J/h$ and $S$. The 1qp sector consists of two 1qp bands $\omega^S(\vec{k},n)$ with $\vec{k}$ the momentum and $n\in\{\rm low,high\}$ the band index due to the two-site unit-cell. The gap $\Delta^S$ is the perturbative minimum of the lower 1qp band $\omega(\vec{k},{\rm low})$ located at momentum $\vec{k}=\vec{0}$ for all $S$. 

Furthermore, we used the pCUT method to calculate the 1qp spectral weight
$A^{S}_{\mathrm{Gap}}\equiv \vert \langle{\rm ref}\vert \mathcal{O}^{\rm eff}_{\mathrm{Gap}}\vert\Delta^S\rangle \vert^2$ of the gap mode $\ket{\Delta^S}$. The observable $\mathcal{O}^{\rm eff}_{\mathrm{Gap}}$ is a Fourier transform of $\mathcal{O}^{\mathrm{eff}}_i$, which results from the unitary transformation of $\mathcal{O}_i=S^z_{i}$, with same momentum and phase factors within the unit-cell of the local observable as the gap mode. We calculated order 8 for $S=1/2$ and order 6 for general $S$. Additional details on the pCUT method and explicit series of all calculated quantities are given in Appendix~\ref{apndx:pcut} for $S\in\{1/2,1\}$ and in Appendix~\ref{apndx:hf-pcut-res} for general spin $S$. 

\section{Results and Discussion} 
\label{sec:phase_diag}
Our major findings are summarized in Fig.~\ref{Fig:phase-diag}, where we show the computed ground-state phase diagram of the antiferromagnetic spin-$S$ Kitaev's honeycomb model in a (1,1,1)-field as a function of $h/J$ and $1/S$. The results are based on TN calculations for $S\in\{1/2,1,3/2,2\}$ and high-order series expansions for general $S$ in the high-field polarized phase. 

The TN results for the pure Kitaev model \mbox{$h=0$} and $S\leq 2$ are shown in Fig.~\ref{fig2}. Our estimated ground-state energy per site, $\varepsilon_0$, obtained from $D=11$ iPEPS simulations with SU algorithm and CTMRG, are $\varepsilon_0\in\{-0.19230, -0.63727, -1.33696, -2.29122\}$ for \mbox{$S\in\{1/2, 1, 3/2, 2\}$}, respectively. Let us stress that these are only qualitative values obtained from simple update and lower energies can indeed be obtained by pushing the simulations to higher bond dimensions or by  using more expensive schemes such as the full tensor update.

Furthermore, the non-vanishing $W_p$ plaquette operator in the Kitaev phase is a unique signature of the gauge ordering around the hexagonal plaquettes of the honeycomb lattice which persist even for larger spin values. We find for all spin values a flux-free ground state with $W_p=1$ and vanishing spin-spin correlations beyond nearest neighbor consistent with earlier findings \cite{Baskaran2007,Baskaran2008b}. Our TN analysis for the pure Kitaev limit also points towards vanishing magnetization for all spin values $S$ (see Fig.~\ref{fig3}-middle panel at $h/J=0$). In addition, for $S=1/2$, the bond energies are fully symmetric on $x$, $y$, and $z$ bonds in agreement with the exact QSL ground state. In contrast, in all other cases ($S>1/2$) the bond energies are anisotropic forming degenerate patterns like  with two of the bonds having the same energies while the third one being considerably different, i.e., forming dimerized patterns on the honeycomb lattice consistent with the semi-classical picture \cite{Rousochatzakis_2018}. The gapped or gapless nature of the ground states cannot be distinguished concretely within iPEPS calculations since the iPEPS is an ansatz for the ground state and any direct information about the energy gap is not available. However, some empirical signatures can be observed in TN simulations, rather by calculating correlation lengths from the eigenvalues of the transfer matrix (not always conclusive) or by observing the convergence speed of the algorithm. It has already been known that the TN simulations converge faster with $1/D$ for gapped phases. We observe that our simulations converge faster for $S>1/2$, suggesting a gapped nature for the QSLs' ground states for $S>1/2$ in agreement with the semi-classical limit \cite{Rousochatzakis_2018} and earlier TN calculations for $S=1$ \cite{Lee2020}. 

\begin{figure}[t]
	\centerline{\includegraphics[width=\columnwidth]{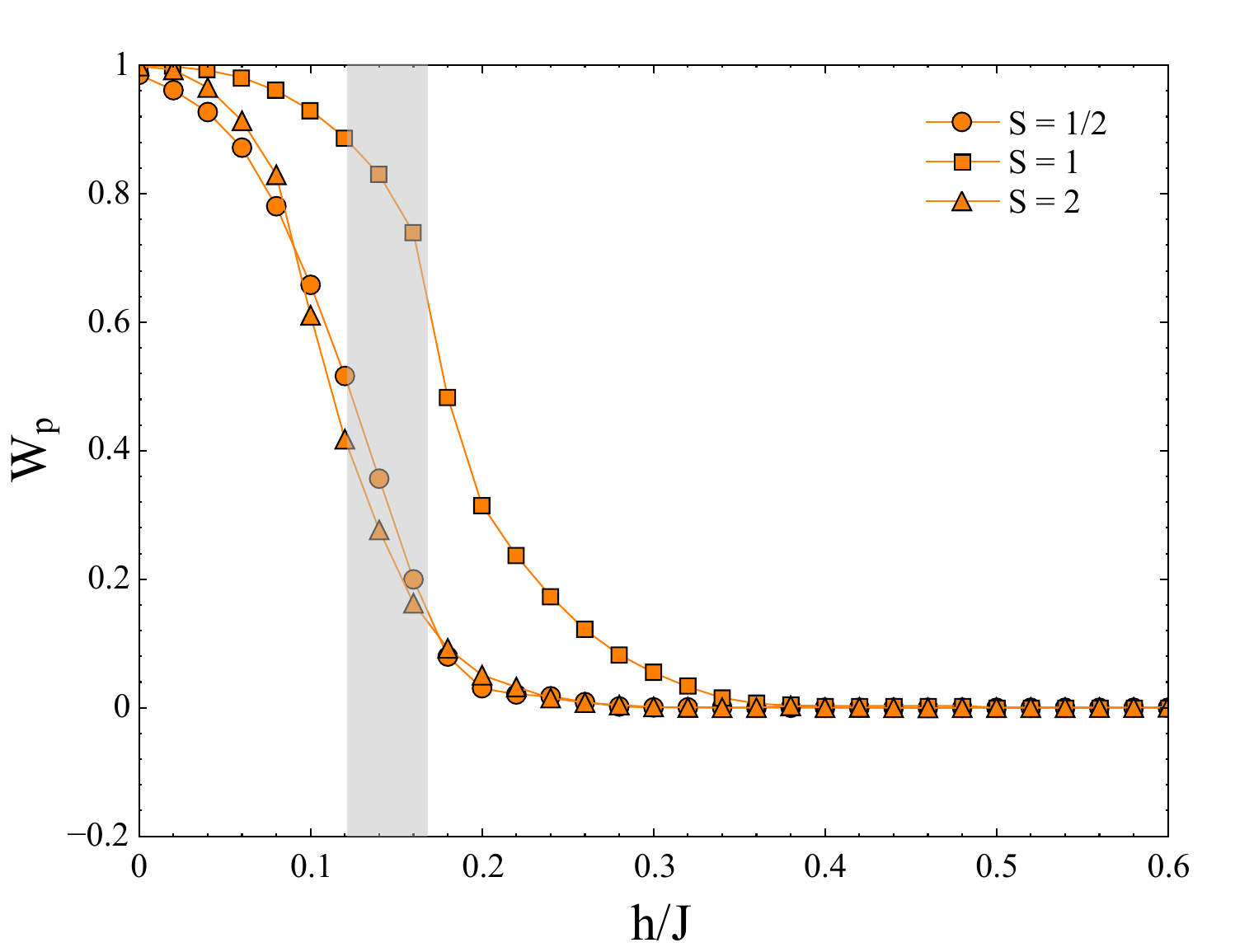}}
	\caption{{\bf $W_p$ plaquette operators as a function of magnetic field for different spin values.} The shaded region denotes the tentative location of the quantum phase transition out of the Kitaev phase for different spins obtained from analysis of other observables such as magnetization and bond energies. By increasing the field strength, the gauge ordering is destabilized in the Kitaev phase and fully vanishes both in the intermediate and polarized phases.}
	\label{Fig:wp}
\end{figure}

\begin{figure*}  
	\centerline{\includegraphics[width=18cm]{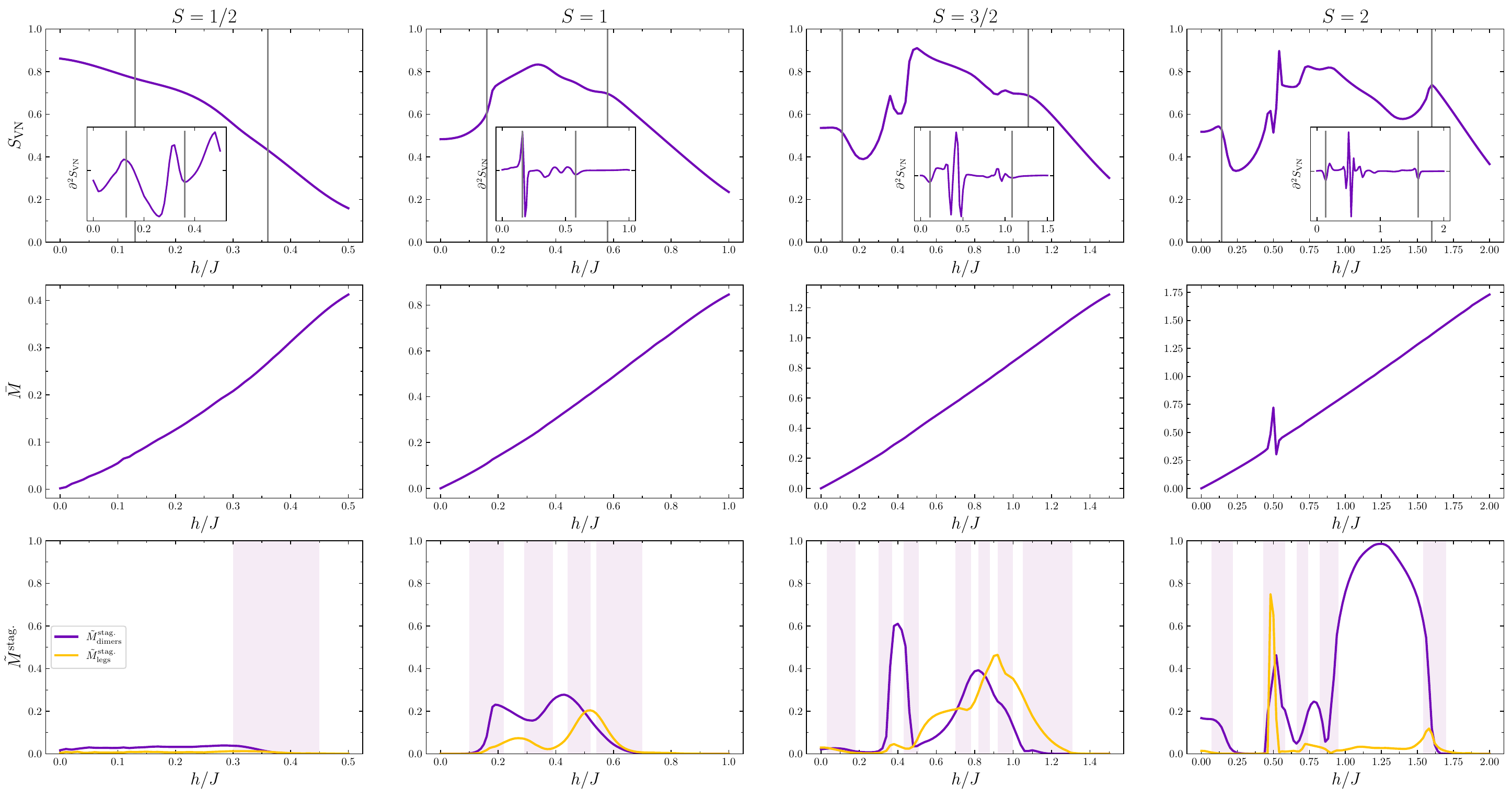}}
	\caption{{\bf Bond entropy and magnetization as a function of the magnetic field.} (Upper panel) Tensor network results for the average bond-entanglement entropy $S_{\rm VN}$ and its second derivative (insets) taken over all bonds as a function of $h/J$ for different spin values. The gray lines delimit the intermediate region from the Kitaev phase for small magnetic fields and the polarized phase for large fields. (Middle panel) Total magnetization as a function of $h/J$ for different spins. (Lower panel) The submagnetization $\tilde{M}_{\text{legs}}^{\text{stag.}}$ and $\tilde{M}_{\text{dimers}}^{\text{stag.}}$ (see Appendix \ref{apndx:sub_mag} for definition) as function of $h/J$ for different spins. The purple shaded regions mark areas of potential phase boundaries.}
		\label{fig3}
\end{figure*}

We now turn in our analysis to small but finite magnetic fields $h/J\lesssim 0.15$. Except for the $S=1/2$ case, where an infinitesimal field opens an energy gap, the ground state for small but finite fields is adiabatically connected to the ground state of the pure Kitaev model and can therefore be characterized the same way. In the phase diagram in Fig.~\ref{Fig:phase-diag} this region is called Kitaev phases and extends for all considered spin values until $h/J\approx 0.15$. Although the expectation value of the $W_p$ plaquette operator is no order parameter of the topological Kitaev phase, it is a characteristic feature of it. Our TN simulations show that the Kitaev phase in the presence of a magnetic field can still be characterized by measuring the $W_p$ plaquette operator which is non-zero in the Kitaev region and strongly reduced in other phases (see Fig.~\ref{Fig:wp}). 
Let us note that the TN simulations for the plaquette operator for $S=3/2$ in the Kitaev region with finite field $0<h\lesssim 0.2$ do not converge properly. This is surprising as in the pure Kitaev limit the TN simulations are well converged to $W_p\approx 1$. Within the accuracy of our results, we cannot fully exclude fundamental physical reasons. In agreement to the finding that the region $h/J\lesssim 0.15$ is adiabatically connected to the pure Kitaev model for all $S>1/2,$ the anisotropic bond energies on the $x$, $y$, and $z$ links form the same degenerate patterns for small fields as for the pure Kitaev model. In order to make sure the anisotropy of bond energies are not a numerical artifact, we performed another set of TN simulations for Kitaev couplings slightly away from the isotropic point, e.g., ($J_x=1.1, J_y=0.95, J_z=0.95$). As expected, we found that the bond with larger exchange coupling always has lower energy (see Appendix~\ref{apndx:tn}-3). Let us further note that similar remarks hold as well for the pure ferromagnetic Kitaev model for all considered spin $S$ (not shown here). 

For $h/J\gtrsim 0.15$, we find an intermediate regime for all $S$ which increases in size when increasing $S$ (see Fig.~\ref{Fig:phase-diag}). To detect potential phases and phase boundaries, we analyzed different quantities obtained from the iPEPS simulations ranging from bond energies  to components of local magnetization $\expectval{S_i^\alpha}$ ($\alpha=x,y,z$) with $i$ the different sites in the unit-cell, and bond-entanglement entropy \mbox{$S_{\rm VN}=-\sum_i \lambda_i^2 \log_2(\lambda_i^2)$} where $\lambda_i$'s are the singular values obtained from local singular-value decomposition in the process of the simple-update. 

Fig.~\ref{fig3}-(upper panel) demonstrates the average bond entropy on all links of the honeycomb lattice. For each $S$, we observe several features in the bond entropy and its second derivative (insets) that may indicate phase boundaries. We mark the first and last extreme point of the second derivative with gray lines to delimit an intermediate region, separating the Kitaev and the polarized phases. The values for these delimiting boundaries are given in Table~\ref{tab:h_c}.

To faithfully capture the potential phases and phase boundaries in these intermediate regions, we further analyze the magnetization on all vertices in the iPEPS unit-cell. Fig.~\ref{fig3}-(middle panel) depicts the average total magnetization $\bar{M} =1/N \sum_i |\langle{\vec{S_i}}\rangle|$, where $N$ is the number of lattice sites in the unit-cell. As expected, the total magnetization starts from zero at $h=0$ within the Kitaev phases and grows by increasing the field strength. Due to its smooth behavior, detecting potential phase boundaries from the average total magnetization curves or their derivatives is not conclusive (there is one notable exception for $S=2$, where a discontinuity occurs at $h/J\approx0.5$). Let us further stress that similar weak features in the second derivative of the ground-state energy have already been observed in ED simulations for $S=1$ \cite{Hickey_2020}.

 To gain further insights, we take a closer look at the magnetization by subtracting the average total magnetization $\bar{M}^{\alpha}$ from the magnetization of each individual site within the iPEPS unit-cell $\tilde{M}_i^{\alpha}=\langle S_i^{\alpha}\rangle - \bar{M}^{\alpha}$ to remove the uniform background. We take the following two approaches:

First, we visualize the components of local magnetization as a function of $h/J$ to identify unique patterns and possible phase boundaries indicated by changes between them for a given iPEPS unit cell. The corresponding magnetization patterns are depicted in Figs.~\ref{Fig:mag_S05}-\ref{Fig:mag_S20} in Appendix~\ref{apndx:visual_mag} accompanied by movie animations that facilitate tracking the evolution of these magnetization patterns as a function of $h/J$. Also note that we represent bond energies by the thickness of the bonds in Figs.~\ref{Fig:mag_S05}-\ref{Fig:mag_S20}. For $S=1/2$ the bond energy is always isotropic. However, for $S>1/2$ the bond energy anisotropies previously reported for the pure Kitaev limit (see Fig.~\ref{fig2}) are visible to increasingly larger field values with increasing $S$ until $h/J\lesssim 0.5$. These bond energy anisotropies even extend to the intermediate region for $S>1$. 

Second, we introduce two types of staggered magnetization $\tilde{M}_{\text{legs}}^{\text{stag.}}$ and $\tilde{M}_{\text{dimers}}^{\text{stag.}}$ (see Fig.~\ref{Fig:sub_mag} in Appendix~\ref{apndx:sub_mag}) aiming to qualitatively characterize the different magnetization patterns (see Fig.~\ref{fig3}-(middle panel)). Within the accuracy of our iPEPS calculations, precisely determining the phase boundaries is challenging due to the smooth changes in these staggered magnetizations. However, we can identify regions where the staggered magnetizations either significantly increase or decrease, or where the two quantities intersect (indicated by purple shaded regions in Fig.~\ref{fig3}). These markers allow us to pinpoint regions that potentially indicate phase transitions. The boundaries identified by both approaches are also reported in Table~\ref{tab:h_c}. Note, we cannot undoubtedly associate all regions with phases spontaneously breaking distinct symmetries since the observed changes in the magnetization patterns may also occur because of a change in the relative phase factors in the ordering momentum. Therefore, not all boundaries are necessarily due to a quantum phase transition but may also be due to crossovers.

\begin{figure*}  
	\centerline{\includegraphics[width=18cm]{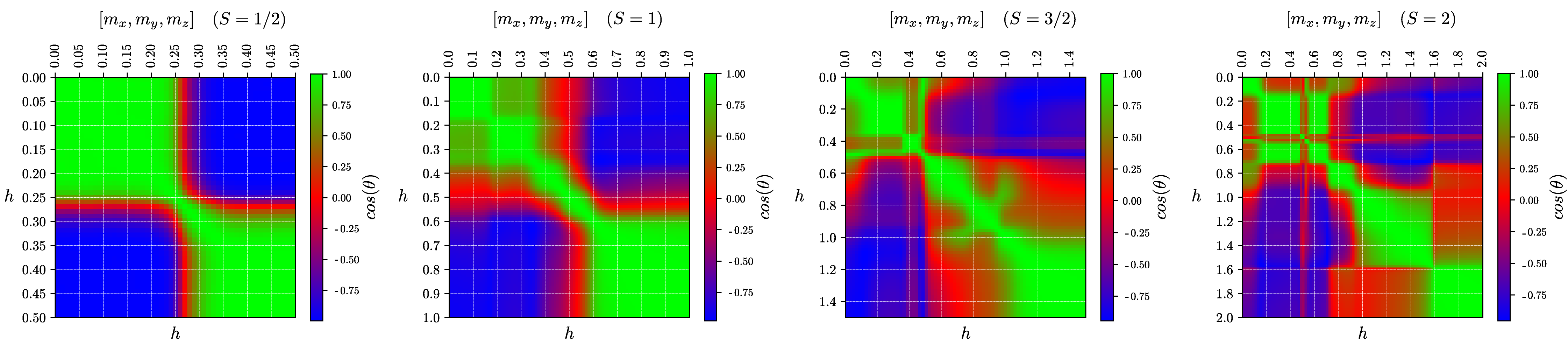}}
	\caption{{\bf Principle component analysis results.} PCA-enhanced similarity maps obtained from local magnetization for different $S$. Each green diagonal block distinguishes a separate phase in the phase diagram. See text for details.}
		\label{Fig:PCA}
\end{figure*}

\begin{figure*}  
	\centerline{\includegraphics[width=18cm]{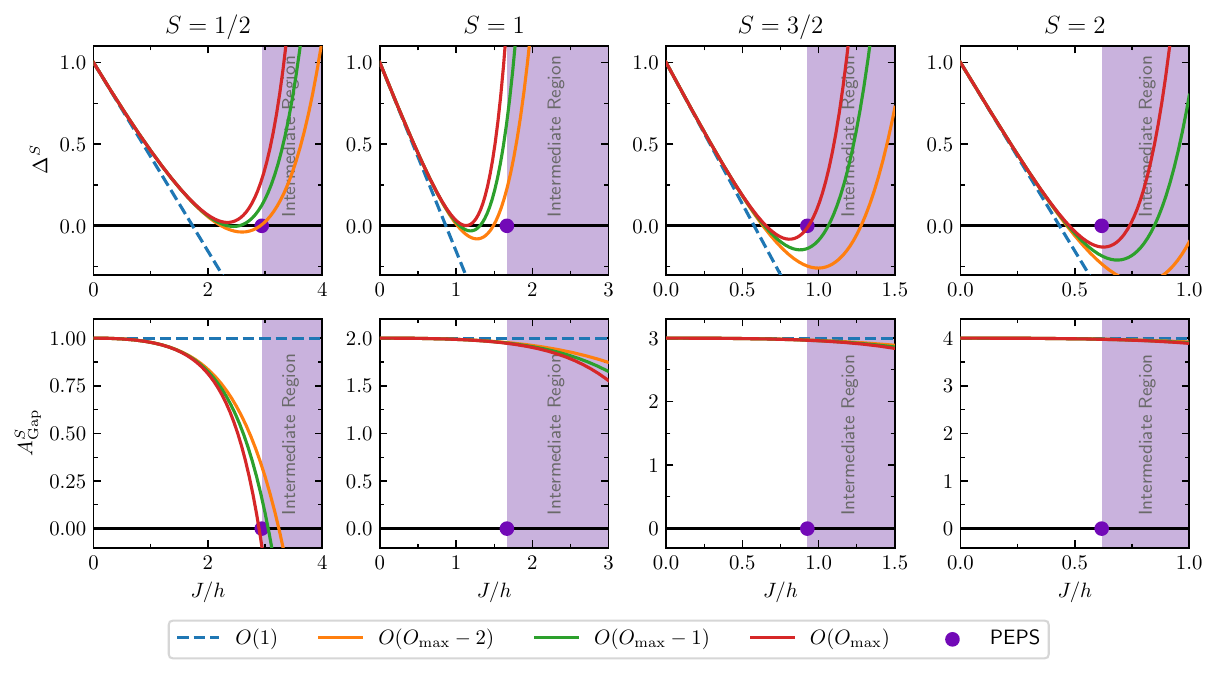}}
	\caption{{\bf Perturbative continuous unitary transformation results for the gap and spectral weight.} (upper panel) pCUT results for the 1qp gap $\Delta^S$ and  (lower panel) the 1qp spectral weight $A^S_{\mathrm{Gap}}$  of the high-field polarized phase as a function of $J/h$ for $S=1/2$ (left), $S=1$ (middle left), $S=3/2$ (middle right), and $S=2$ (right). Dashed lines correspond to the first-order results while solid lines represent the three highest perturbative orders for each quantity ($O_{\rm max}$ denotes the maximal order reached). Purple circles and shaded regions illustrates the phase transition points and the intermediate regions as obtained from the iPEPS calculations.}
		\label{fig4}
\end{figure*}

To provide a more rigorous insight into the phase boundaries, we further employ principal component analysis (PCA) \cite{gewers_principal_2021,shlens_tutorial_2014} of the local magnetization data. Initially, we create feature vectors by concatenating the components of on-site magnetization, denoted as $m_i^\alpha$ ($\alpha=x,y,z$), for all vertices of the lattice under different fields. Subsequently, we generate a 2D similarity map by calculating the dot product of feature vectors for all field values. To enhance the resolution of the similarity map, we apply PCA analysis to the similarity data (see Appendix~\ref{apndx:pca} for PCA details). Fig.~\ref{Fig:PCA} illustrates the PCA-enhanced similarity maps for different $S$. Each square block along the diagonal of the color map characterizes a single (phase) region in the phase diagram. The location of the phase boundaries between two adjacent phases can be further detected at the intersecting corners of the neighboring blocks. The phase boundaries for different $S$ from the PCA analysis are reported in Table~\ref{tab:h_c}. 

As pointed out above, the intermediate phase for \mbox{$S=1/2$} has been predicted to be a gapless spin liquid. Locally, the Kitaev and the field induced intermediate QSL phases of the $S=1/2$ model are indistinguishable. This is also evident in the PCA analysis as the two QSL phases are distinguished as a single large block in the similarity map of Fig.~\ref{Fig:PCA}. The second lower block further corresponds to the polarized phase. We note that the phase boundary between these two blocks has a significant uncertainty (see range given in Table~\ref{tab:h_c}). As a consequence, we plot the transition points for $S=1/2$ in Fig.~\ref{Fig:phase-diag} from the derivatives of the bond entropy.

\begin{table*}
	\begin{tabular}{cc||cccccccc}
	\hline
	$\mathbf{S=1/2}$ &  bond ent. &    0.17     &  -- & 0.34   &&&&& \\
	&  ani. &    -- & 0.25(?)     &  0.41      &&&&& \\
	&  submag. &     --     & -- & (0.3, 
	0.45) &&&&& \\
	&  pca  &    -- & -- & (0.24, 0.32) &&&&& \\ \hline\hline
	$\mathbf{S=1}$    &  bond ent. &   0.16     &    --   &     --     & 0.59 &&&&   \\
	&  ani. &   0.14     &    0.355    & 0.455, 0.505 & 0.7    &&&&   \\
	&  submag. & (0.1, 0.18) & (0.29, 0.39) & (0.44, 0.52) &   (0.54, 0.7)   &  &&& \\
	&  pca &   0.16     &    0.35     &     0.48     & 0.60 &&&&   \\ \hline\hline
	$\mathbf{S=3/2}$  & bond ent. &     0.11   &    --  &  --  &     --   &     --        &    --    &      1.08   &     \\
	& ani. &     0.15    &     0.35    &  0.47, 0.49  &     0.75(?)   &     --        &    0.97    &      1.28   &     \\
	& submag. & (0.03, 0.18) & (0.30, 0.37) & (0.43, 0.51) & (0.7, 0.78) & (0.82, 0.88) & (0.92, 1.0) & (1.05, 1.31) & \\
	& pca &     0.1     &     0.35    &     0.49     &    --        &       --      &    0.96    &    1.08     &    \\ \hline\hline        
	$\mathbf{S=2}$   & bond ent. &    0.14     &       --    &     --     &     --    &    1.60 &&&\\
	& ani. &    0.15     &    0.47, 0.51, 0.55    &     0.71     &     0.93    &    1.66 &&& \\
	& submag. & (0.07, 0.22) & (0.43, 0.57) & (0.66, 0.74) & (0.82, 0.95) & (1.54, 1.7) &&& \\	
	& pca &    0.12     &       0.54    &     0.70     &     0.92    &    1.62 &&&\\ \hline
\end{tabular}
\caption{Summary of the analysis of the intermediate region for all spin values $S$ showing all potential phase transitions. The phase boundaries from the analysis of the bond-entanglement entropy (bond ent.), from visualizing the magnetization by animation (ani.), from the submagnetization plots (submag.), and from the PCA analysis (pca). The boundaries from the submagnetization plots are denoted by brackets and (?) denote boundaries where we were unsure if the magnetization pattern is really distinct from the  previous one. For $S=1/2$ the PCA analysis can not distinguish between the QSL states in the Kitaev and intermediate regions and the values were obtained from bond entropy (see text for more details).} 
\label{tab:h_c}
\end{table*}

We summarize our results from the analysis of the bond-entanglement entropy, magnetization animations, submagnetization and PCA analysis in Table~\ref{tab:h_c}. First, note that for $S=1/2$ our analysis of the magnetization is blind to the different QSL phases. Yet, we can detect the phase boundaries of the intermediate QSL phase from the entanglement entropy in agreement with previous results. For $S>1/2$, the bond-entanglement entropy and the analysis of the magnetization give consistent estimates pinpointing an intermediate region in-between the Kitaev phases for small magnetic field and the high-field polarized phase for large field values. In this region we observe non-vanishing, distinct magnetization patterns which indicates the presence of several possible intermediate phases with distinct local ordering from the point of view of magnetization and bond energies.

 Finally, we discuss the breakdown of the high-field polarized phase towards the intermediate region. Our corresponding results for the 1qp gap $\Delta^S$ and the 1qp spectral weight $A^S_{\mathrm{Gap}}$ are shown in Fig.~\ref{fig4}. We expect the gap $\Delta^S$ to close at the quantum critical point $J_{\rm crit}$. We note, however, that a potential first-order phase transition for $J<J_{\rm crit}$ can not be detected with the gap series, but is absent in our TN calculations. Also, existing DMRG data \cite{Gohlke_2018,Zhu_2018,Zhu_2020} points towards a continuous phase transition. For increasing $S$, the gap closes for smaller values of $J_{\rm crit}$ so that the polarized phase is reduced (see also Fig.~\ref{Fig:phase-diag}). In the limit $S\rightarrow\infty$ one obtains $2 S J_{\rm crit} = \sqrt{3}$. Interestingly, the breakdown of the polarized phase is found at rather large values of $J_{\rm crit}$ by TN and DMRG calculations implying that $\Delta^S$ closes very flat as a function of $J$, i.e.,~$\Delta^S\propto (J-J_{\rm crit})^{z\nu}$ with critical exponent $z\nu$ substantially larger than one. For $S=1/2$, this can be also seen in the dynamic structure factor calculated by DMRG \cite{Gohlke_2018}. This peculiar behavior is likely also the reason why conventional extrapolation techniques like Pad\'{e} and Dlog Pad\'{e} approximation \cite{Guttmann1989} do not work for $\Delta^S$ (see Appendix~\ref{apndx:hf-pcut-res} and Ref.~\cite{Adelhardt2017} for similar findings) and we restrict the discussion mostly to the bare series. The 1qp spectral weight $A^S_{\mathrm{Gap}}$ behaves opposite to the gap in the sense that higher perturbative orders become less important compared to the zeroth order $2S$. As a consequence, while our series data is consistent for a vanishing 1qp spectral weight at $J_{\rm crit}$ for $S=1/2$, the spectral weight remains finite in the whole polarized phase for $S>1/2$. Moreover, the ratio $\Delta^S/A^S_{\mathrm{Gap}}$ allows a consistent extrapolation for all $S$. Critical points obtained from Pad\'{e} and Dlog Pad\'{e} approximants are displayed in Fig.~\ref{Fig:phase-diag} as solid phase transition lines and show convincing agreement with our TN results. Overall, also our series-expansion results display the difference between $S=1/2$ and $S>1/2$ consistent with the findings from TN simulations.

\section{Conclusions and outlook} 
\label{sec:conclude}
We applied TN approaches and linked-cluster expansions to investigate the ground-state phase diagram of the antiferromagnetic Kitaev's honeycomb model in a (1,1,1)-field for general spin values. While an intermediate region between the low-field Kitaev and the high-field polarized phase is found for all studied $S$ consistent with the literature, our results show distinct behavior for $S=1/2$ and $S>1/2$. For the Kitaev phases, our TN calculations confirm the absence of fluxes and spin-spin correlations beyond nearest neighbor in the ground state, but signal a breaking of the discrete orientational symmetry for $S\in\{1,3/2,2\}$ inline with the semiclassical limit \cite{Rousochatzakis_2018} which can be clearly seen in the anisotropies of bond energies. Further, the TN simulations hint towards gapped Kitaev phases for all $S>1/2$.

We further identified an intermediate region with vanishing plaquette operator ($W_p\approx 0$) for all spin values. For $S=1/2$ our results are in line with the presence of a QSL phase in this regime, for $S>1/2$ we observed a sequence of subregions with distinct, non-vanishing magnetization patterns and anisotropic bond energies for each S. Analyzing our results with various approaches, we could consistently pinpoint potential phase boundaries suggesting that the regions correspond to different low-energy phases. However, we cannot fully exclude possible crossovers occurring between symmetry-related magnetization patterns.We further found that the number of  intermediate (phase) regions increases with $S$ so that the ground-state phase diagram of the antiferromagnetic Kitaev's honeycomb model in a (1,1,1)-field  becomes more and more complex.

 For the high-field polarized phase, the previous numerical indications of a gap closing for $S\in\{1/2,1\}$ \cite{Gohlke_2018,Hickey2019,Hickey_2020} and the analysis of our perturbative series of the excitation gap at zero momentum suggest an unconventional quantum critical breakdown for all $S$. This is consistent with the presence of exotic physics at intermediate Kitaev couplings. Further, in line with our iPEPS findings, a difference between $S=1/2$ and $S>1/2$ can be seen in the associated spectral weights. 

Antiferromagnetic Kitaev models in finite magnetic fields represent therefore an exciting and very challenging playground for general values of $S$. A deeper understanding of the intermediate region as well as of the quantum critical behavior is certainly still needed. This includes also the generalization to arbitrary field directions and anisotropic Kitaev couplings.

\section{Additional Information}
\subsection{Acknowledgments}
We thank S. Trebst for fruitful discussion. M. Gohlke and F. Pollmann for providing DMRG data. The TN simulations were performed on ATLAS HPC cluster at DIPC. KPS acknowledges financial support by the German Science Foundation (DFG) through the grant SCHM 2511/11-1. KPS further acknowledges financial support by the German Science Foundation (DFG) through the grant SCHM 2511/11-1 and the TRR 306 QuCoLiMa ("Quantum Cooperativity of Light and Matter") - Project-ID 429529648- as well as the Munich Quantum Valley (MQV) including the MQV lighthouse project QuMeCo ("Quantum Measurement and Control for the Enablement of Quantum Computing and Quantum Sensing"), which is supported by the Bavarian state government with funds from the Hightech Agenda Bayern Plus.

RO and SSJ acknowledge financial support from Ikerbasque and DIPC. 

\subsection{Author Contributions}
Tensor Network simulations were done by Saeed Jahromi and Roman Orus. Series Expansion calculations were done by Max H\"ormann, Sebastian Fey, Patrick Adelhardt and Kai Phillip Schmidt. Max H\"ormann performed the white-graph expansions for general $S$. PCA analysis was performed by Hooman Karamnejad. Patrick Adelhardt performed data analysis of the magnetization. All authors contributed to the interpretation of data and the manuscript.

\subsection{Competing Interests}
The authors declare no competing interests.

\subsection{Data Availability}
Correspondence and requests for data should be addressed to Saeed Jahromi. 

\subsection{Code Availability}
Correspondence and requests for code should be addressed to Saeed Jahromi.

\bibliography{bibliography.bib}
\bibliographystyle{apsrev4-1}

\newpage
\onecolumngrid
\appendix
\begin{figure}
\centerline{\includegraphics[width=10cm]{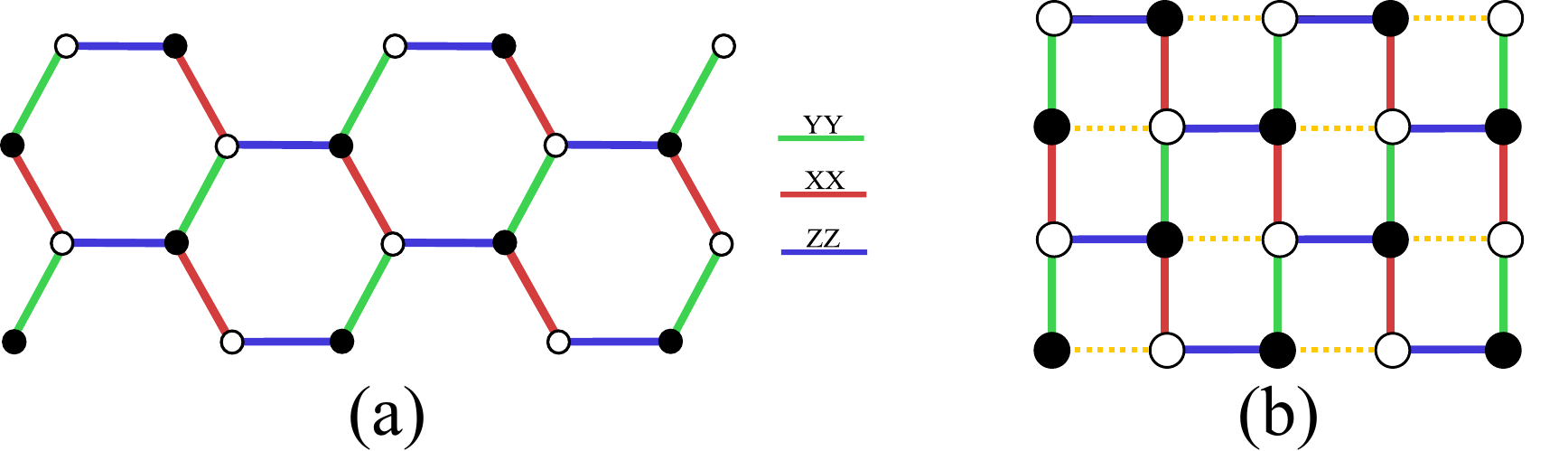}}
\caption{{\bf The translationally invariant Honeycomb lattice.} (a) The Kitev model on the honeycomb lattice. (b) Reshaping the honeycomb lattice (a) into the brick-wall structure. By adding dummy links to the lattice, the resulting brick-wall structure will be turned into a square lattice.}
\label{Fig:brickwall}
\end{figure}

\section{Tensor-Network Algorithms in more details}
\label{apndx:tn}
In this section, we provide the details of the tensor-network (TN) algorithms that have been used for simulating the spin-$S$ Kitaev model on the honeycomb lattice.

\subsection{The iPEPS method for the honeycomb lattice}
Exact contraction of an infinite 2D TN is a computationally hard problem. One therefore has to resort to approximation techniques such as tensor renormalization group (TRG) \cite{Levin2007,Gu2008} or corner transfer matrix renormalization group (CTMRG) \cite{Nishino1996,Orus2009,Corboz2014a,Corboz2010a}. The CTMRG have been proven to be one the most efficient and accurate techniques for contraction of TN with square geometries. Therefore, current state-of-the-art iPEPS algorithms are usually developed for square lattices. In order to simulate other 2D lattice with iPEPS method, one then needs to map the underlying lattice to a square structure. To this end, one can coarse-grain several lattice sites into a square lattice of block-sites \cite{Jahromi2018} or fine-grain \cite{Schmoll2020} the lattice by locally splitting both lattice sites and the corresponding Hilbert space to obtain a square geometry.   

\begin{figure}
\centerline{\includegraphics[width=10cm]{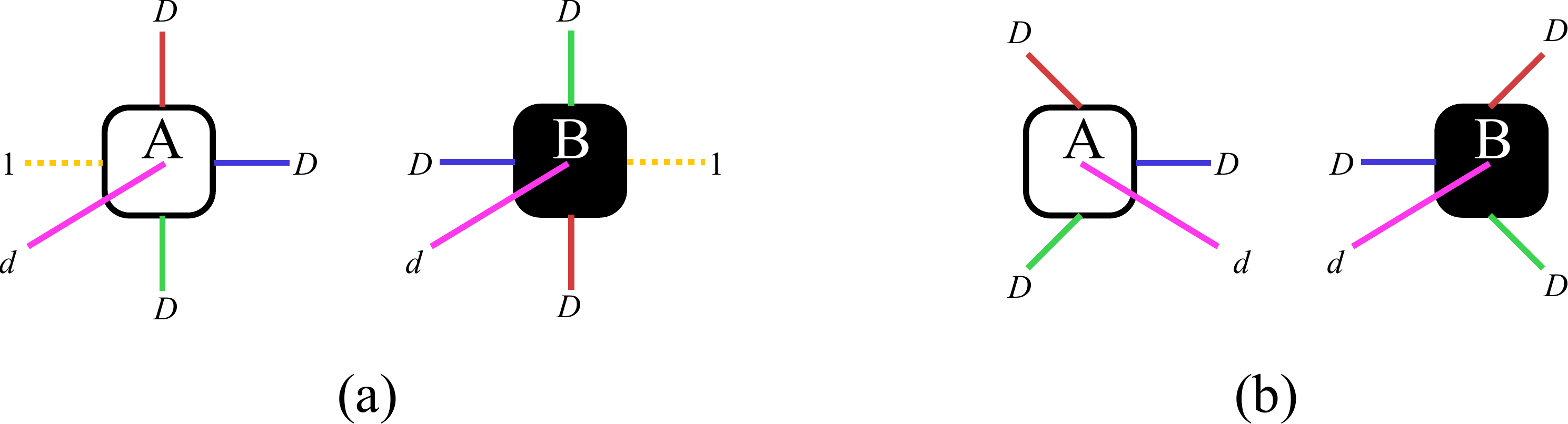}}
\caption{{\bf The iPEPS local tensors.} (a) The rank-5 PEPS tensors which are obtained by reshaping the honeycomb lattice into a brick-wall structure and adding dummy indices with dimension $D=1$ to the tensors. (b) The rank-4 tensors on the original honeycomb lattice which are used in the gPEPS algorithm.}
\label{Fig:tensors}
\end{figure}

Fortunately, the honeycomb lattice already has a rectangular geometry when reshaped into the brick-wall structure. One can then add a dummy link to each site of the lattice to obtain a full square lattice. This mapping is shown in Fig.~\ref{Fig:brickwall} for the Kitaev model on the honeycomb lattice. The corresponding square TN can further be obtained by assigning rank-5 tensors to each site of the lattice given the virtual dimension of the dummy link be set to $D=1$, as shown in Fig.~\ref{Fig:tensors}-(a). The resulting TN will then have a square geometry and can be efficiently simulated by the iPEPS code already developed for the square lattice. The only difference is that no Hamiltonian term acts on the trivial index of the two neighboring tensors (see dashed yellow legs in Fig.~\ref{Fig:tensors}-(a)).

Our iPEPS techniques for simple update is based on the iterative evolution of the Kitaev Hamiltonian in imaginary time. The renormalization and truncation of the local tensors are performed by singular-value decomposition. The full environment used in the calculation of the expectation values of local operators have been approximated by the CTMRG algorithm.

Within our iPEPS simulations, we managed to push the simulations up to bond dimension $D=11$ for SU. The boundary dimension $\chi$ for approximating the environment in the CTMRG was at least $\chi=D^2$. However, we always checked to make sure we are fully converged. Besides, we fixed $\chi=64$ for $D>8$. 

\subsection{The gPEPS method for any infinite lattice}
The gPEPS technique \cite{Jahromi2019}, in contrast to the standard iPEPS method which is mostly developed for the square geometry, can be adapted to any lattice geometry and any spatial dimension. The gPEPS algorithm targets the geometrical challenges for TN simulation of local Hamiltonians by resorting to the so-called {\it structure-matrix} (SM), first proposed by some of the authors in Ref.~\cite{Jahromi2019}. Each column of the SM corresponds to one of the links of the lattice and contains all the details about the neighboring tensors, their interconnecting indices, and their bond dimensions. Thanks to this, one can fully automatize the TN update by looping over the columns of the SM in a very systematic way, without the burden of complications due to geometry (see Refs.\cite{Jahromi2019} for detailed discussions). 

The gPEPS algorithm uses an optimized version of the SU algorithm for optimizing the TN and further uses a mean-field-like environment for calculating the expectation values of local operators, hence allowing for simulating a larger bond dimension $D$. Similar to the iPEPS, the gPEPS algorithm is also based on the imaginary-time evolution of the Hamiltonian. The main difference here is that in contrast to the iPEPS in which we mapped the honeycomb lattice to a square TN of rank-5 tensors with one dummy index (Fig.~\ref{Fig:tensors}-(a)), the gPEPS considers rank-4 tensors (Fig.~\ref{Fig:tensors}-(b)) in a translationally invariant unit-cell directly on the original honeycomb lattice. The connectivity information of the honeycomb unit-cell used in our gPEPS calculations has been shown for a $8$-site unit-cell in Table ~\ref{tab:smhoneycomb}.

\begin{table}[h]
\caption{Structure-matrix of a translationally invariant honeycomb lattice with $8$-site unit-cell. The physical dimension of each tensor $T_i$ is labeled by zero. The virtual indices therefore are started from one in each row of the table.}
\centering
\begin{tabular}{l|cccccccccccc}
  \hline
 & $E_1$ & $E_2$ & $E_3$ & $E_4$ & $E_5$  & $E_6$ & $E_7$ & $E_8$ & $E_9$ & $E_{10}$ & $E_{11}$ & $E_{12}$  \\ 
  \hline
$T_1$ & 1 & 2 & 3 & 0 & 0 & 0 & 0 & 0 & 0 & 0 & 0 & 0 \\
$T_2$ & 1 & 0 & 0 & 2 & 3 & 0 & 0 & 0 & 0 & 0 & 0 & 0 \\
$T_3$ & 0 & 0 & 0 & 1 & 0 & 2 & 3 & 0 & 0 & 0 & 0 & 0 \\
$T_4$ & 0 & 1 & 0 & 0 & 0 & 2 & 0 & 3 & 0 & 0 & 0 & 0 \\
$T_5$ & 0 & 0 & 0 & 0 & 1 & 0 & 0 & 0 & 2 & 3 & 0 & 0 \\
$T_6$ & 0 & 0 & 1 & 0 & 0 & 0 & 0 & 0 & 2 & 0 & 3 & 0 \\
$T_7$ & 0 & 0 & 0 & 0 & 0 & 0 & 0 & 1 & 0 & 0 & 2 & 3 \\
$T_8$ & 0 & 0 & 0 & 0 & 0 & 0 & 1 & 0 & 0 & 2 & 0 & 3 \\
		  \hline 
\end{tabular}
\label{tab:smhoneycomb} 
\end{table}

We were able to push simulations of the spin-$S$ Kitaev model in the presence of a uniform magnetic field up to $D=18$ for different $S$ values. The phase boundaries were further detected by calculating different local quantities such as energy and its derivatives, magnetization, nearest-neighbor spin correlation, and Von Neumann bond entanglement entropy as shown in the main text.

\begin{figure*}  
	\centerline{\includegraphics[width=18cm]{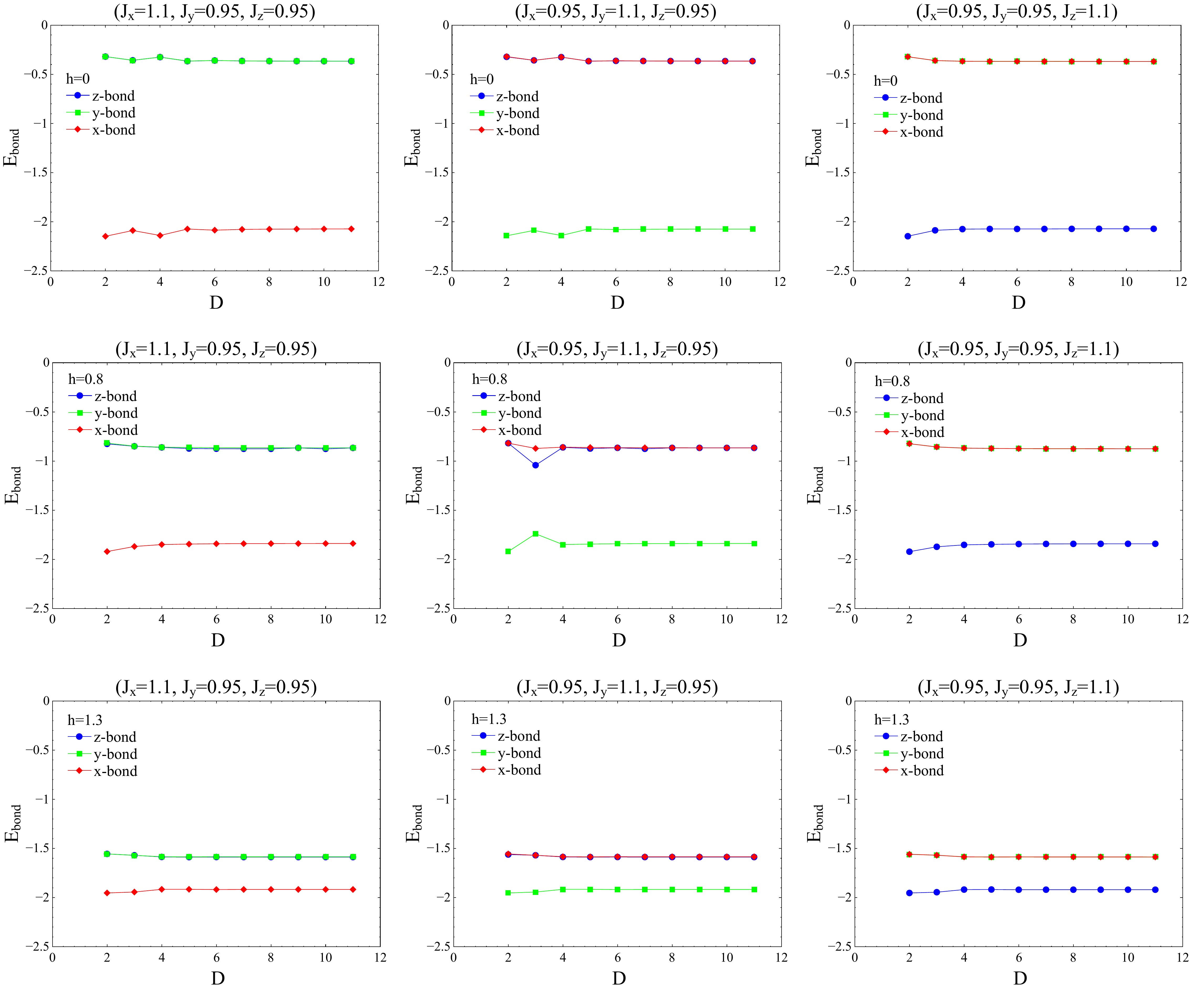}}
	\caption{{\bf Ground-state energy per-site of the spin-$3/2$ Kitaev model for different field strength and anisotropic exchange coupling.} The bond with larger exchange coupling always has lower energy. As we approach the polarized phase, the bond energies on $x$-, $y$- and $z$-links tend to approach each other until they eventually become fully identical in the polarized phase. Similar results holds for any $S>1/2$.}
		\label{fig_anisotropic}
\end{figure*}

\subsection{Tensor-network results for the anisotropic Kitaev model}
In order to provide a more clear picture about the nature of different phases in the phase diagram of the antiferromagnetic Kitaev model in a magnetic field, here we present our complimentary results for the anisotropic Kitaev regime, e.g., ($J_x =1.1, J_y=095, J_z=0.95$). Similar to the isotropic case ($J_x=J_y=J_z$), we observe that by moving slightly away from the isotropic point the bond energies on the $x$, $y$, and $z$-links for $S>1/2$ become anisotropic and form degenerate patterns with two of the bonds having the same energies while the third one being considerably different, forming dimerized patterns on the honeycomb lattice which are consistent with the semi-classical limit \cite{Rousochatzakis_2018}. Fig.~\ref{fig_anisotropic} demonstrates the bond energies of the spin-$3/2$ antiferromagnetic Kitaev model for several field strength $h$ and different exchange anisotropies. One can clearly see that the bond with larger exchange coupling always has lower energy. As we approach the polarized phase by increasing the magnetic field, the bond energies on $x$-, $y$- and $z$-bond tend to approach each other until they eventually become fully identical in the polarized phase. Similar results holds for any $S>1/2$.

\section{Visualization of magnetization data}
\label{apndx:visual_mag}

\begin{figure}
	\centerline{\includegraphics[width=10cm]{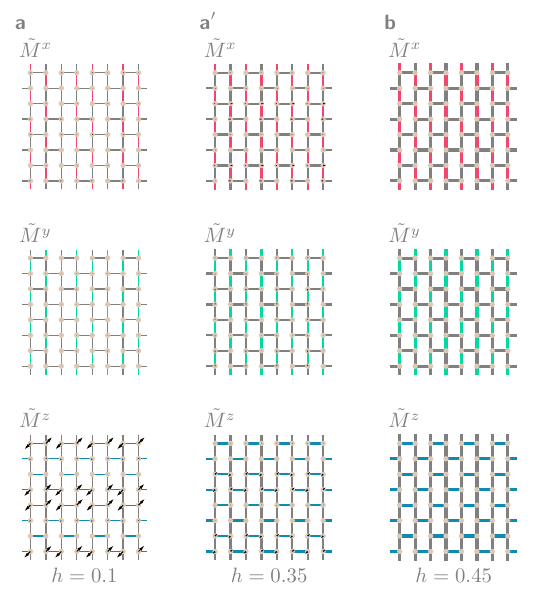}}
	\caption{{\bf Visualization of the magnetization in $\tilde{M}^x$, $\tilde{M}^y$, and $\tilde{M}^z$ (arrows) on Kitaev honeycomb lattice for $S=1/2$ based on the unit-cell of the iPEPS calculations.} Red, green, and blue bonds indicate $x$-, $y$-, and $z$-interactions respectively. The thickness of each bond is proportional to the bond energy. The magnetization is visualized for different magnetic field values $h$, chosen such that there is exactly one representative for each phase. The phases are also numerated by distinct letters.}
	\label{Fig:mag_S05}
\end{figure}

\begin{figure}
	\centerline{\includegraphics[width=10cm]{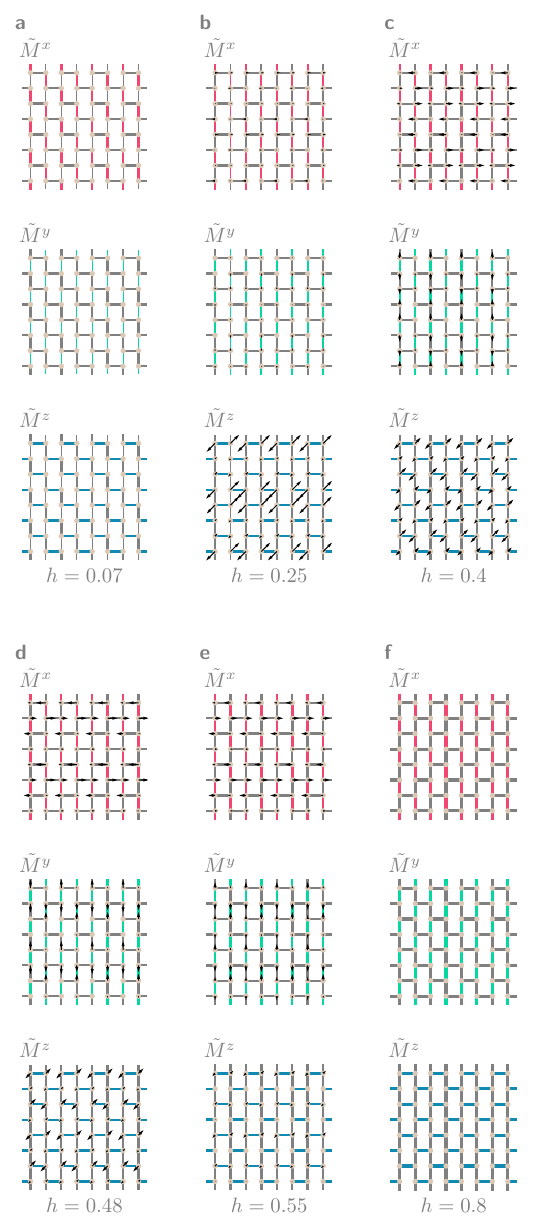}}
	\caption{{\bf Visualization of the magnetization in $\tilde{M}^x$, $\tilde{M}^y$, and $\tilde{M}^z$ (arrows) on Kitaev honeycomb lattice for $S=1$ based on the unit-cell of the iPEPS calculations.} Red, green, and blue bonds indicate $x$-, $y$-, and $z$-interactions respectively. The thickness of each bond is proportional to the bond energy. The magnetization is visualized for different magnetic field values $h$, chosen such that there is exactly one representative for each phase. The phases are also numerated by distinct letters.}
	\label{Fig:mag_S10}
\end{figure}

\begin{figure}
	\centerline{\includegraphics[width=13cm]{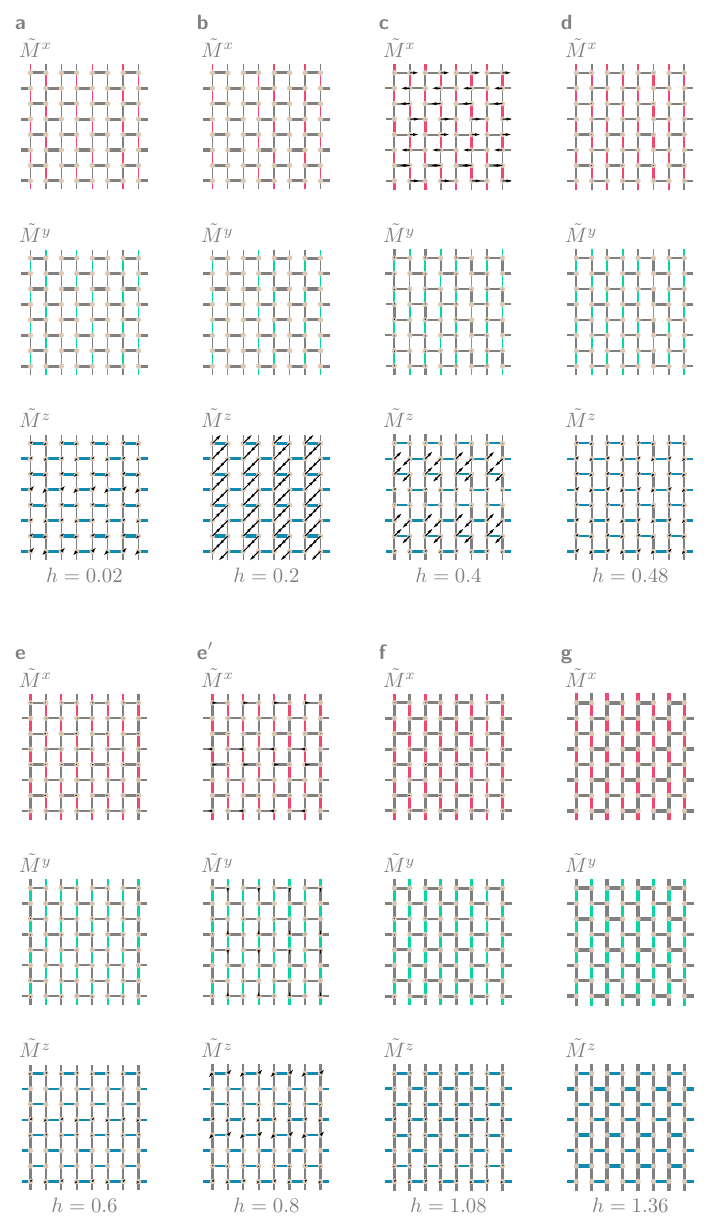}}
	\caption{{\bf Visualization of the magnetization in $\tilde{M}^x$, $\tilde{M}^y$, and $\tilde{M}^z$ (arrows) on Kitaev honeycomb lattice for $S=3/2$ based on the unit-cell of the iPEPS calculations.} Red, green, and blue bonds indicate $x$-, $y$-, and $z$-interactions respectively. The thickness of each bond is proportional to the bond energy. The magnetization is visualized for different magnetic field values $h$, chosen such that there is exactly one representative for each phase. The phases are also numerated by distinct letters.}
	\label{Fig:mag_S15}
\end{figure}

\begin{figure}
	\centerline{\includegraphics[width=13cm]{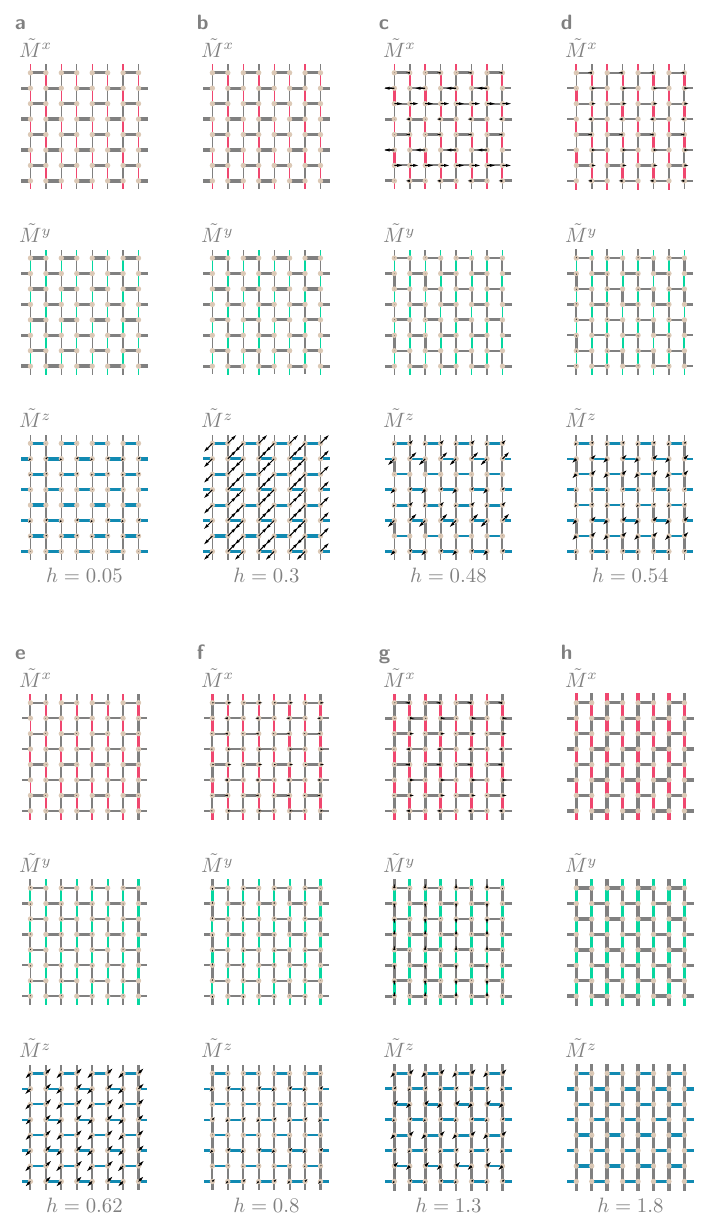}}
	\caption{{\bf Visualization of the magnetization in $\tilde{M}^x$, $\tilde{M}^y$, and $\tilde{M}^z$ (arrows) on Kitaev honeycomb lattice for $S=2$ based on the unit-cell of the iPEPS calculations.} Red, green, and blue bonds indicate $x$-, $y$-, and $z$-interactions respectively. The thickness of each bond is proportional to the bond energy. The magnetization is visualized for different magnetic field values $h$, chosen such that there is exactly one representative for each phase. The phases are also numerated by distinct letters.}
	\label{Fig:mag_S20}
\end{figure}

The total magnetization increases linearly with the magnetic field $h$ (see Fig.~\ref{fig3}-middle panel) and does not show any features signaling a quantum phase transition nor do its derivatives. There is one notable discontinuity for $S=2$ that could indeed be indicative of a quantum phase transition. To shed more light into the intermediate region located between the Kitaev phases and the high-field polarized phase, we visualize the local magnetization on the lattice. The total magnetization is $\bar{M} = |\vec{\bar{M}}|$ with $\vec{\bar{M}} = (\bar{M}^x, \bar{M}^y, \bar{M}^z)^T$ and where the magnetization for each spin component $\alpha\in\{x,y,z\}$ is defined as
\begin{equation}
	\bar{M}^{\alpha} =\frac{1}{N} \sum_{i} \expectval{S_i^{\alpha}}.
\end{equation}
Here, $N$ is the number of sites in the unit cell. The total magnetization increases smoothly with the magnetic field $h$ and can be seen as a background which can be subtracted from the local magnetization per site as
\begin{equation}
\tilde{M}_i^{\alpha} = \expectval{S^{\alpha}_i} - \bar{M}^{\alpha}
\end{equation}
to make possible magnetization patterns on the honeycomb lattice more visible. We plot the local magnetization per site $\tilde{M}_i^{\alpha}$ as arrows for each individual component $\alpha\in\{x,y,z\}$ on the honeycomb lattice (in brick-wall structure) for every magnetic field value $h$ and animate the plots in a movie. The direction of the arrows is determined by the sign of the local magnetization component and the length of the arrows reflects the magnitude. For visualization purposes the length of arrows is not absolute for different spin values but rather relative to the maximal magnetization value for each spin. Further, we visualize the energy of each bond by the line thickness that is proportional to the bond energy. The movie animations can be found as additional supplemental material to this publication. With these animations we are able to identify distinct magnetic patterns by eye to determine possible quantum phases and phase boundaries between them. In Figs.~\ref{Fig:mag_S05}-\ref{Fig:mag_S20} we show each magnetization pattern for a fixed field value $h$ representative of each potential quantum phase. From just looking at the figures it is hard to undoubtedly identify unique patterns that are truly distinct and not symmetry related.

\newpage
\newpage
\newpage

\section{Defintion of (sub)magnetization regimes}
\label{apndx:sub_mag}

\begin{figure}
	\centerline{\includegraphics[width=0.5\textwidth]{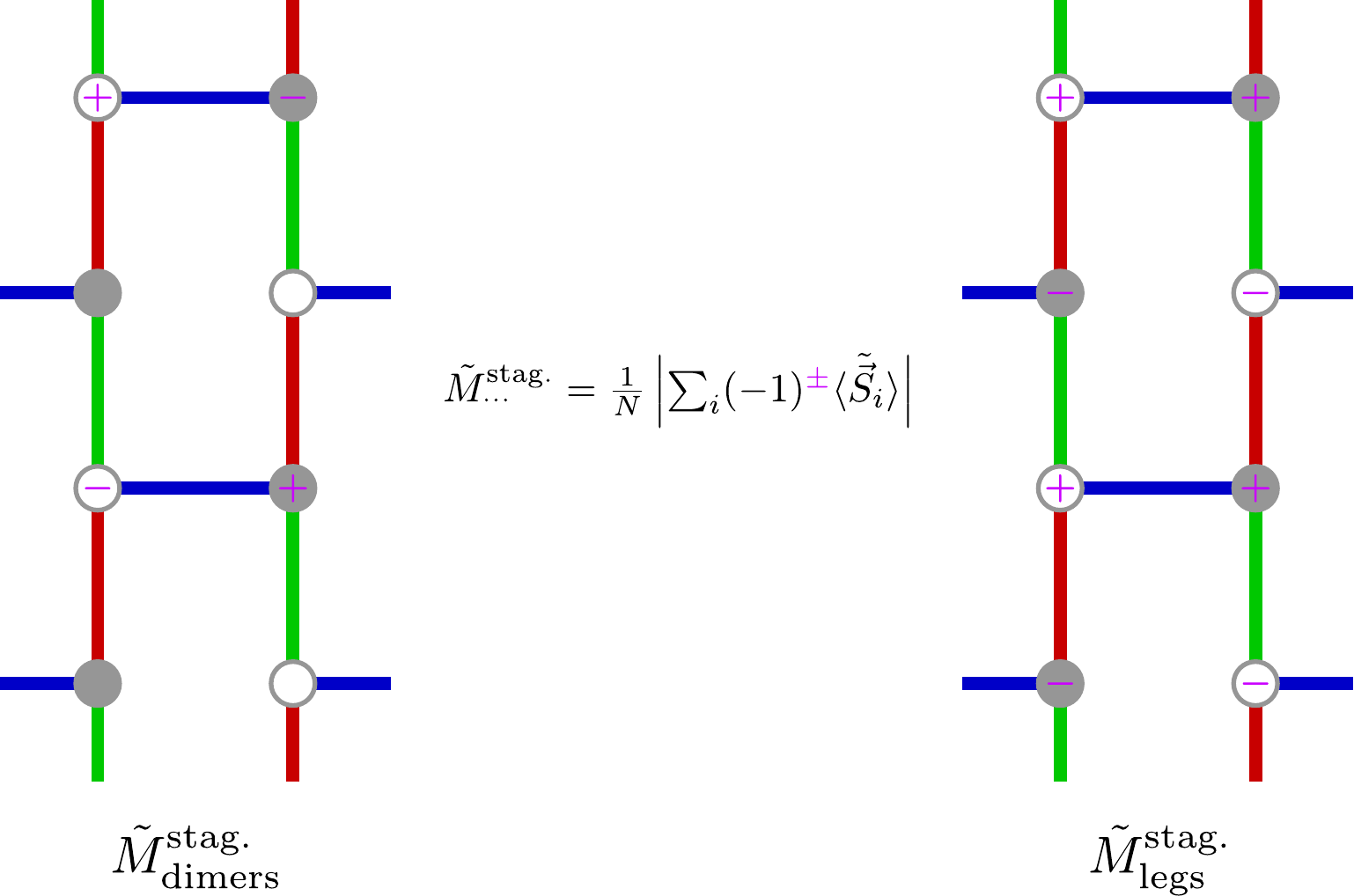}}
	\caption{{\bf Definition of the staggered magnetizations according to the unit-cell.} Illustration of the iPEPS unit cell depicting the signs associated with the sites (pink plus and minus signs) for the definition of the staggered magnetizations $\tilde{M}^{\text{stag.}}_{\text{dimers}}$ and $\tilde{M}^{\text{stag.}}_{\text{legs}}$. The bond colors red, green, and blue are associated with $x$-, $y$-, and $z$-interactions.} 
	\label{Fig:sub_mag}
\end{figure}

Another way to qualitatively identify the magnetization patterns and their phase boundaries already found by visualizing the magnetization per site on the lattice is to define submagnetization quantities. Again, the average magnetization on the iPEPS unit-cell is given by
\begin{equation}
	\bar{M}^{\alpha} =\frac{1}{N}_{\text{uc}} \sum_{i} \expectval{S_i^{\alpha}}
\end{equation}
for the spin component $\alpha\in\{x,y,z\}$. With $\vec{\bar{M}} = (\bar{M}^x, \bar{M}^y, \bar{M}^z)^T$ we again subtract the total magnetization from the local magnetization per site as $\vec{\tilde{M}}_i = \expectval{\vec{S}_i} - \vec{\bar{M}}$. Thus, the staggered magnetization for $\vec{\tilde{M}}_i^{\alpha} $ can be written as
\begin{equation}
	\tilde{M}^{\text{stag.}}_{\cdots} = \frac{1}{N} \left\vert \sum_{i} (-1)^{\pm} \vec{\tilde{M}}_i \right\vert.
\end{equation}
We define two types of staggered (sub-)magnetization on the dimers $\tilde{M}^{\text{stag.}}_{\text{dimers}}$ and along the legs $\tilde{M}^{\text{stag.}}_{\text{legs}}$ as depicted in Fig.~\ref{Fig:sub_mag} for the iPEPS unit-cell, where we depict the sign associated with each site. These two quantities are plotted against the magnetic field in Fig~\ref{fig3}-(lower panel).

\section{PCA analysis}
\label{apndx:pca}

\begin{figure}[h]
\centerline{\includegraphics[width=0.3\textwidth]{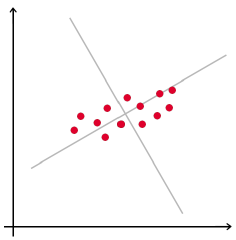}}
\caption{{\bf Principal component axis.} Mapping of the data space on orthogonal directions.}
\label{Fig:pca-axis}
\end{figure}

Principal Component Analysis (PCA) is a prevalent technique that reduces the dimensionality of multidimensional data while preserving its essential features \cite{gewers_principal_2021,shlens_tutorial_2014}. This method effectively distills critical information from complex datasets. PCA operates by mapping data into orthogonal directions known as principal components. It prioritizes components based on their variance, identifying those that encapsulate the most significant information within the dataset. In other words, PCA analysis tries to identify new coordinates that offer the most informative representation of data. 

Consider a dataset as plotted in Fig.~\ref{Fig:pca-axis}, where each axis represents a projection of the data. The central challenge involves determining two optimal axes (depicted as gray coordinates in the figure) such that the projection of the data points onto these axes maximizes the variance along each respective axis. The new representation enhances the interpretability and utility of the data by focusing on these more informative axes. The principal components can be found through the following steps as detailed below:

\begin{enumerate}
    \item {\bf Normalization:} The data must first be standardized such that all the variables are transformed into the same scale
\[ z = \frac{{\rm value} - {\rm mean}}{{\rm standard} \enskip {\rm deviation}} \]
    
    \item {\bf Covariance matrix:} The covariance of the normalized data must be computed to ascertain the relationships between the variables within the dataset and to understand how they deviate from the mean relative to each other. This step is essential for identifying the extent to which variables are interdependent, providing insight into the underlying structure of the data.
\newline
\[\begin{bmatrix}
	{\rm cov}(x_1,x_1) & \cdots & {\rm cov}(x_1,x_m)\\
	\vdots & \ddots & \vdots \\
	{\rm cov}(x_m,x_1) & \cdots & {\rm cov}(x_m,x_m)\\
\end{bmatrix}\]

    \item {\bf Eigenvalue Decomposition of the Covariance matrix:} To identify the principal components of the data, it is necessary to compute the eigenvalues and eigenvectors of the covariance matrix. For instance, consider a two-dimensional dataset with variables x and y. The principal components can be discerned by examining the eigenvalues, $\lambda_i$, and eigenvectors, $v_i$, of the covariance matrix.   
\newline
\[ v_1 = \begin{bmatrix}
	a_1 \\
	b_1 \\
\end{bmatrix} \quad \lambda_1 = c_1\]
\newline
\[ v_2 = \begin{bmatrix}
	a_2 \\
	b_2 \\
\end{bmatrix} \quad \lambda_2 = c_2\]
\newline
If $ \lambda_1 > \lambda_2 $, our first principal components is $v_1$ and the second is $v_2$ and vice versa. For a multidimensional dataset, we can choose to keep the first \textit{K} principal components by sorting the eigenvectors according to their eigenvalues. The new dimension of the dataset is then represented by \textit{K}. 
\end{enumerate}

In order to adopt the above PCA analysis to the tensor network data of the Kitaev-Field model, we must first create the feature vectors by concatenating the components of local magnetization on all sites of the lattice. The feature vector must be created for all the field values $h$.
\[ \forall h \in \left[ h_{\rm min}, h_{\rm max} \right] \Rightarrow
\bf{X} = \left[ \bf{m_x}, \bf{m_y}, \bf{m_z} \right] \]

Next we apply the PCA analysis to the feature vectors and obtain a refined representation for magnetization vectors in terms of the principle components. Subsequently, we generate a 2D similarity map by calculating the dot product of the refined feature vectors for all field values. As discussed in the main text, the similarity map enables us to detect the phase boundaries more accurately as compared to the analysis of the bare magnetization data.

\section{The pCUT Method}
\label{apndx:pcut}
The pCUT method maps the Hamiltonian 
\begin{align}
\mathcal{H}_{\rm KIF}=J\sum_{{\alpha\text{-links} \atop \alpha=x,y,z }\atop <i,j>}S_i^\alpha S_j^\alpha+h\sum\limits_{i}\left(S_i^x+S_i^y+S_i^z\right) \equiv \mathcal{V} + \mathcal{H}_0, \label{eq:kitaev_hamiltonian_pcut}
\end{align} unitarily to an effective Hamiltonian $\mathcal{H}_{\rm eff}$, which conserves the number of quasi-particles in the polarized high-field phase where spins point in $(1,1,1)$-direction. This mapping is done perturbatively up to high orders in powers of $x\equiv J/h$. The quasi-particles in the polarized phase correspond to dressed spin-flip excitations, which are adiabatically connected to localized spin flips above the fully polarized state in the limit $h\rightarrow\infty$.

It is convenient to perform a local unitary transformation in the spin-$S$ Hamiltonian \eqref{eq:kitaev_hamiltonian_pcut}
to a basis, where the field term $\mathcal{H}_0$ is diagonal. The unperturbed part then reads 

\begin{equation}
\mathcal{H}_0 = \sqrt{3}h \sum_i \tilde{S}_i^{z},
\end{equation}
while for the perturbation $\mathcal{V}$ the expression becomes 

\begin{equation}
\mathcal{V} = J/6 \sum_{{\gamma\text{-links} \atop \alpha, \beta, \gamma =x,y,z }\atop <i,j>} C^{\alpha\beta}_\gamma \tilde{S}^{\alpha}_i\tilde{S}^{\beta}_j.
\end{equation} 

The coefficients $C^{\alpha\beta}$ are given by 

\begin{align}
C^{\alpha\beta}_x=\left(
\begin{array}{ccc}
1 & \sqrt{3} & -\sqrt{2} \\
\sqrt{3} & 3 & -\sqrt{6} \\
-\sqrt{2} & -\sqrt{6} & 2
\end{array}
\right), \quad
C^{\alpha\beta}_y=\left(
\begin{array}{ccc}
1 & -\sqrt{3} & -\sqrt{2} \\
-\sqrt{3} & 3 & \sqrt{6} \\
-\sqrt{2} & \sqrt{6} & 2
\end{array}
\right) , \quad
C^{\alpha\beta}_z=\left(
\begin{array}{ccc}
4 & 0 & 2 \sqrt{2} \\
0 & 0 & 0 \\
2 \sqrt{2} & 0 & 2
\end{array}
\right).
\end{align}

The unperturbed part $\mathcal{H}_0$ has an equidistant spectrum and can therefore be used as a starting point for pCUT.

The action of the perturbation depends on the link-type $x$, $y$ or $z$ but the difference between the three cases only enters in different phase factors. This can be seen from the form of the $C$-matrices.

We made use of this property to efficiently implement pCUT. For that we employed the white-graph method \cite{Coester2015, Adelhardt2020} where we calculated each contribution from a multi-parameter perturbation sequence on graphs tracking all possible phases of the virtual fluctuations. Only after the pCUT calculation was preformed we attributed the right phases, depending on the link types of the graph embedding on the lattice.

\subsection{Ground-state energy and gap}
We calculated the ground-state energy and gap up to order $9$ in the parameter $x=J/h$ for $S=1/2$ and up to order $8$ for $S=1$ in units of $\sqrt{3}h$. The series are given by

\begin{equation}
\begin{aligned}
\Delta^{S=1/2} &= 0.00000959406432895\,x^9 + 0.0000212393084160\,x^8 + 0.0000414176411990\,x^7 \\ 
& + 0.0000704746622861\,x^6 
+ 0.000237867694925\,x^5 + 0.00111454046639\,x^4 \\
& + 0.00801875373874\,x^3 + 0.0277777777777\,x^2  - 0.577350269189\,x + 1.0 \quad ,
\end{aligned}
\label{Gap_S12}
\end{equation}

\begin{equation}
\begin{aligned}
\Delta^{S=1} &= 0.0101022918713\,x^8 + 0.0115628640647\,x^7 + 0.0141008200463\,x^6 + 0.0194831516594\,x^5 \\
& + 0.0287208504801\,x^4 + 0.0481125224324\,x^3 + 0.0555555555555\,x^2 - 1.15470053837\,x + 1.0 \quad ,
\end{aligned}
\label{Gap_S1}
\end{equation}

\begin{equation}
\begin{aligned}
\varepsilon_0^{S=1/2} &= - 0.00000149464455105\,x^9  - 0.00000367780870199\,x^8 - 0.00000870428476930\,x^7 \\
& - 0.0000231117034186\,x^6  - 0.0000869246303917\,x^5 - 0.000436205353843\,x^4\\
& - 0.00231481481481\,x^3 - 0.0120281306081\,x^2 + 0.125\,x - 0.866025403784 \quad ,
\end{aligned}
\label{GS_S12}
\end{equation}

and
\begin{equation}
\begin{aligned}
\varepsilon_0^{S=1} &= - 0.00294477723047\,x^8 - 0.00390784623983\,x^7 - 0.00543115959436\,x^6\\
& - 0.00806374790428\,x^5 - 0.0129933509655\,x^4 - 0.0231481481481\,x^3\\
& - 0.0481125224324\,x^2 + 0.5\,x - 1.73205080756 \quad .
\end{aligned}
\label{GS_S1}
\end{equation}

\subsection{Observable}

Furthermore we have calculated the spectral weight of the gap mode $\ket{\Delta^S}$ measured by the Fourier transform of the observable $\mathcal{O}_i\equiv S^{z}_{i}$, $\mathcal{O}_{\mathrm{Gap}} = \frac{1}{\sqrt{2\mathcal{N}_\mathrm{u}}}\sum_{i,j}\sum_{\alpha=0}^{1} (-1)^\alpha \mathcal{O}_{i,j,\alpha}$, which reads

\begin{equation}
A^{S}_{\mathrm{Gap}}\equiv\vert \bra{\mathrm{ref}} \mathcal{O}^{\mathrm{eff}}_{\mathrm{Gap}}\ket{\Delta^S}\vert ^2\quad .
\end{equation}
Here $i,j$ represent the coordinate of the unit-cell, $\alpha$ the position in the unit-cell and $\mathcal{N}_\mathrm{u}$ is the number of unit-cells. The observable thus has the same phase factors as the gap mode. The unitary transformation of pCUT is used to transform $\mathcal{O}_{\mathrm{Gap}}$ to $\mathcal{O}^{\mathrm{eff}}_{\mathrm{Gap}}$. For $S=1/2$ it is given up to order $8$ by 

\begin{equation}
\begin{aligned}
A^{S=1/2}_{\mathrm{Gap}} & = - 0.0000447670776607\,x^8 - 0.0000874502157678\,x^7 - 0.000206177711284\,x^6 \\
& - 0.000531879498960\,x^5 - 0.00220407521719\,x^4 - 0.00534583582582\,x^3 - 0.0138888888888\,x^2 + 1.0\quad .
\end{aligned}
\label{Obs_S12}
\end{equation}

\section{High-field pCUT expansion for general spin $S$}
\label{apndx:hf-pcut-res}
\subsection{Ground-state energy and gap}

The goal is to obtain the ground-state energy per site $\varepsilon_0$ and the gap $\Delta$ as a function of $x=J/h$ for a general spin value $S$ in units of $\sqrt{3}h$. While for fixed spin values this is performed regularly with the pCUT method, deducing linked-cluster expansion for general $S$ has not been done to the best of our knowledge. Consequently, we will describe this part in more detail in the following. For that purpose it is instructive to write down the general action of $\tilde{S}^x$- and $\tilde{S}^y$-operators in the eigenbasis $\ket{m}$ with $m\in\{-S,\ldots,+S\}$ of the $\tilde{S}^z$-operator,

\begin{eqnarray}
\bra{m'} \tilde{S}^x \ket{m} &=&\frac{1}{2}\sqrt{S(S+1)-m'm}\; \left(\delta_{m',m+1}+\delta_{m'+1,m}\right)\nonumber\\
\bra{m'} \tilde{S}^y \ket{m} &=&\frac{1}{2{\rm i}}\sqrt{S(S+1)-m'm}\; \left(\delta_{m',m+1}-\delta_{m'+1,m}\right)\nonumber\\
\bra{m'}\tilde{S}^z \ket{m} &=& m\; \delta_{m',m}\quad .\label{eq::S_tilde}
\end{eqnarray}

In the following we describe our approach for the ground-state energy $\varepsilon_0$. Contributions in the perturbative expansion correspond to virtual fluctuations. In the case of $\varepsilon_0$ the amplitudes of these fluctuations are given by expectation values with respect to the reference state $\ket{-S-S\ldots -S}$. In order $k$ perturbation theory the expectation values then involve operator sequences containing $k$-times the perturbation $\mathcal{V}$. Obviously, for terms in $\mathcal{V}$ containing solely spin operators $\tilde{S}^x$ and $\tilde{S}^y$, one only gets a finite contribution to $\varepsilon_0$ in order $k$ perturbation theory if each site has been acted on an even number of times. As a consequence, the final contribution will be a polynomial in $S$ of order $k$, since the square-root expressions in Eq.~\eqref{eq::S_tilde} for matrix elements from $m$ to $m'$ will always occur with an even multiplicity. Further, for the term proportional to $\tilde{S}^z\tilde{S}^z$, one has to note that the constant contribution $\propto S^2$ is not part of the perturbation $\mathcal{V}$ since it is just a constant times the identity matrix. The leading term is therefore proportional to $S$. Finally, for $\tilde{S}^x\tilde{S}^z$, the contribution proportional to $(-S)\sqrt{S}$ is zero because of destructive interferences of the three Kitaev links on each vertex. Here the leading term is proportional to $\sqrt{S}$. Altogether, each term in $\mathcal{V}$  scales at most with $S$ so that in order $k$ perturbation theory one obtains a polynomial of order $k$ in $S$.

Thus one way to calculate the general $S$ contribution to the ground-state energy $\varepsilon_0$ is to perform the calculation for a sufficiently large number of $S$-values allowing to determine the polynomial in $S$ exactly. Instead of doing this we applied a more efficient approach described in the following. We introduce three perturbation parameters to distinguish contributions of order $S$, $\sqrt{S}$, and $1$. This way only a single calculation has to be performed. Further, the maximal local Hilbert space needed is only of dimension $\lfloor k/2\rfloor$ in contrast to $k+1$ that one needs for the naive approach. We then exploit that 
\begin{eqnarray}
\sqrt{S(S+1)-(-S+d)(-S+d+1)}^2 &=&2(d+1)S-d(d+1)\nonumber\\
                               &=&(\sqrt{2(d+1)S}-\sqrt{d(d+1)})(\sqrt{2(d+1)S}+\sqrt{d(d+1)})\quad .
\end{eqnarray}
Whenever a spin operator $\tilde{S}^x$ or $\tilde{S}^y$ acts and increases the eigenvalue of $\tilde{S}^z$, we assign to the corresponding spin operator the factor $(\sqrt{2(d+1)S}+\sqrt{d(d+1)})$. In contrast, if the eigenvalue of $\tilde{S}_z$ decreases, we track the factor $(\sqrt{2(d+1)S}-\sqrt{d(d+1)})$. This way each action of $\mathcal{V}$ can be decomposed into contributions proportional to $S$, $\sqrt{S}$, and $1$.  This explains how $\tilde{S}^x\tilde{S}^x$-, $\tilde{S}^x\tilde{S}^y$- and $\tilde{S}^y\tilde{S}^y$-processes have to be treated. All other terms can be handled in a similar manner straightforwardly.

In practice, we have calculated the ground-state energy for general values of $S$ up to order 7 in $x$. Up to order $4$, this expression reads

\begin{equation}
\begin{aligned}
 \varepsilon_0^S &= (- 0.0191559117092\,S^4 + 0.00623680846346\,S^3 - 0.0000742477198031\,S^2)\,x^4\\
 & + (- 0.0277777777777\,S^3 + 0.00462962962962\,S^2)\,x^3 -0.0481125224324\,S^2\,x^2\\
 & + 0.5\,S^2\,x - 1.73205080756\,S \quad .
\end{aligned}
\end{equation}

Note that the general expression agrees with the expressions in Eq. \eqref{GS_S12} and Eq. \eqref{GS_S1} for $S=1/2$ and $S=1$. Further, for $S\in\{\frac{3}{2}, 2\}$, the series expansions read

\begin{equation}
\begin{aligned}	
 \varepsilon_0^{S=3/2} &= - 0.0949621005193\,x^7 - 0.0829288164821\,x^6 - 0.0766192986968\,x^5\\
 & - 0.0760946318332\,x^4 - 0.0833333333333\,x^3 - 0.108253175473\,x^2\\
 &  + 1.125\,x - 2.59807621135
 	 \\	
 \varepsilon_0^{S=2} &= - 0.841028333911\,x^7 - 0.533302274710\,x^6 - 0.357243560432\,x^5\\
 & - 0.256897110519\,x^4 - 0.203703703703\,x^3 - 0.192450089729\,x^2\\
 & + 2.0\,x - 3.46410161513
 	\quad .
 \end{aligned}
\end{equation}

The same approach can also be realized for the one-particle gap $\Delta$ and we also have reached order 7 for general values of $S$. Up to order $4$, this expression reads

\begin{equation}
\begin{aligned}
 \Delta^S &= (0.00437242798353\,S - 0.0329218106995\,S^2 + 0.0572702331961\,S^3)\,x^4\\
 & +(0.0641500299099\,S^2 - 0.0160375074774\,S)\,x^3 + 0.0555555555555\,S\,x^2\\
 &-1.15470053837\,S\,x + 1 \quad .
 \end{aligned}
\end{equation}

The general expression here also agrees with the expressions in Eq. \eqref{Gap_S12} and Eq. \eqref{Gap_S1} for $S=1/2$ and $S=1$. Further, for $S\in\{1,\frac{3}{2}, 2\}$, the series expansions read  

\begin{equation}
\begin{aligned}
 \Delta^{S=1} &= 0.0101022918713\,x^8 + 0.0115628640647\,x^7 + 0.0141008200463\,x^6 + 0.0194831516594\,x^5 \\
 & + 0.0287208504801\,x^4 + 0.0481125224324\,x^3 + 0.0555555555555\,x^2 - 1.15470053837\,x + 1.0
 \\	
 \Delta^{S=3/2} &= 0.217999872570\,x^7 + 0.170528114759\,x^6 + 0.143725023609\,x^5 + 0.125771604938x^4\\
 & + 0.120281306081\,x^3 + 0.0833333333333\,x^2 - 1.73205080756\,x + 1.0
 	 \\	
 \Delta^{S=2} &= 1.57595243639\,x^7 + 0.895840698332\,x^6 + 0.539783672884\,x^5 + 0.335219478737\,x^4 \\
 & + 0.224525104684\,x^3 + 0.111111111111\,x^2 - 2.30940107675\,x + 1.0
 	\quad .
 \end{aligned}
\end{equation}

\subsection{Dispersion}

Using the full graph decomposition we do not only have information about the energy gap but also about the complete one-particle spectrum. There are two atoms per unit-cell which yields two bands after diagonalization via Fourier transform. The two dispersions in second order in $x$ for general $S$ are

\begin{equation}
	\begin{aligned}
	& 1 - S/\sqrt{3}\,x + (4/27)((S^2\cos(k_x - k_y))/8 - (3S)/8 - (3S^2)/8 + (S^2\cos(k_x))/8 + (S^2\cos(k_y))/8)\,x^2 \\
	&\pm S\,x(2x/(3\sqrt{3})-2)/(6\sqrt{3})\sqrt{2\cos(k_x-k_y)+2\cos(k_x)+2\cos(k_y)+3}.
	\end{aligned}
\end{equation}

\subsection{Observable}
In principle the calculation of the local observable $\mathcal{O}_i\equiv S^{z}_{i}$ for general $S$ is performed by the same technique used for the ground-state energy and the one-quasi-particle gap described above. 

Special care has to be taken when implementing the action of the observable. One again has to distinguish between observable processes that increase or decrease the $\tilde{S}^z$-value.

This means, e.g., for the $\tilde{S}^x$-part of $S^z$ one has to distinguish behaviour where $\tilde{S}^x$ increases the $\tilde{S}^z$-value by one and attribute $(\sqrt{2(d+1)S}+\sqrt{d(d+1)})$ to it while for processes where the $\tilde{S}^z$-value gets decreased by one one has to use $(\sqrt{2(d+1)S}-\sqrt{d(d+1)})$ instead.

We have calculated the effective observable $\mathcal{O}^{\rm eff}_i$ for general values of $S$ up to order 6 in $x=J/h$. Up to order $4$, this expression reads

\begin{equation}
\begin{aligned}
 A^{S}_{\mathrm{Gap}} & = (0.00167181069958\,S^2 + 0.0155178326474\,S^3 - 0.0729881115683\,S^4)\,x^4 \\
 & + (0.0106916716516\,S^2-0.0641500299099\,S^3)\,x^3 -0.0555555555555\,S^2\,x^2 + 2.0\,S \quad .
\end{aligned} 	
\end{equation}

The general expression agrees with the expression in Eq. \eqref{Obs_S12} for $S=1/2$. Further, for $S\in\{1,\frac{3}{2}, 2\}$, the series expansions read  

\begin{equation}
\begin{aligned}
 A^{S=1}_{\mathrm{Gap}} &= - 0.000131804332821\,x^6 - 0.000393123245477\,x^5 - 0.00122465773873\,x^4\\
 & - 0.00304831580551\,x^3 - 0.00823045267489\,x^2 + 2.0
  \\	
A^{S=3/2}_{\mathrm{Gap}} &= - 0.00191804537762\,x^6 - 0.00359256457652\,x^5 - 0.00687776253619\,x^4\\
& - 0.0109739368998\,x^3 - 0.0185185185185\,x^2 + 3.0
	 \\	
 A^{S=2}_{\mathrm{Gap}} &= - 0.0119676532972\,x^6 - 0.0163624232067\,x^5 - 0.0227595035635\,x^4 \\
 & - 0.0268251790885\,x^3 - 0.0329218106995\,x^2 + 4.0
 	\quad .
\end{aligned}
\end{equation}

\section{Extrapolation}
\label{apndx:pcut-extrapl}
 Extrapolation techniques like Pad\'{e} and DLog Pad\'{e} approximation are the most common techniques to extrapolate high-order series. For a general introduction we refer to Ref.~\cite{Guttmann1989}. The  Pad\'{e} approximant of order $[L, M]$ is given by 
	\begin{align}
		P[L, M] = \frac{P_L(x)}{Q_M(x)} = \frac{p_0 + p_1 x + \cdots + p_L x^L}{1 + q_1 x + \cdots + q_M x^M}
	\end{align}
	where the coefficients $p_i$ and $q_i$ are given uniquely by the requirement that the Taylor series of $P[L,M]$ in the expansion parameter $x$ up to order $L+M$ must correspond to the calculated series $F$ of the same order. Further, DLog Pad\'{e} approximants $dP[L,M]$ correspond to a Pad\'{e} approximation of the logarithmic derivative
	\begin{align}
		D(x) = \frac{\text{d}}{\text{d} x} \ln(F(x))\quad .	
	\end{align}

\end{document}